\documentclass[prb,showpacs,twocolumn,superscriptaddress,floatfix,longbibliography,nofootinbib]{revtex4-2}
\usepackage{graphicx,amsfonts,amssymb,amsmath,xspace,dsfont,bm,bbm}
\usepackage[colorlinks=true,citecolor=green,linkcolor=red,urlcolor=blue]{hyperref}
\usepackage{subfigure}
\usepackage{blindtext}
\usepackage{tabularx}

\usepackage{color}
\definecolor{g}{rgb}{.1,0.4,.1} % {.0,0.7,.5}
\definecolor{b}{rgb}{0,0.2,1}
\definecolor{rouge}{rgb}{0.82,0.,0.}
\definecolor{vert}{rgb}{0.,0.82,0.}
\definecolor{orange}{rgb}{1,0.5,0.}
\definecolor{bleu}{rgb}{0.,0.,0.82}
\definecolor{m}{rgb}{0.82,0.,0.82}
\definecolor{vert2}{rgb}{0.,0.5,0.}
\definecolor{rougeclair}{rgb}{1.0,0.7,0.7}

\newcolumntype{Y}{>{\centering\arraybackslash}X}

\begin{document}

\title{Effective models for dense vortex lattices in the Kitaev honeycomb model}

\author{David J. Alspaugh}
\affiliation{Sorbonne Universit{\'e}, CNRS, Laboratoire de Physique Th{\'e}orique de la Mati{\`e}re Condens{\'e}e, LPTMC, F-75005 Paris, France}

\author{Jean-No{\"e}l Fuchs}
\affiliation{Sorbonne Universit{\'e}, CNRS, Laboratoire de Physique Th{\'e}orique de la Mati{\`e}re Condens{\'e}e, LPTMC, F-75005 Paris, France}

\author{Anna Ritz-Zwilling}
\affiliation{Sorbonne Universit{\'e}, CNRS, Laboratoire de Physique Th{\'e}orique de la Mati{\`e}re Condens{\'e}e, LPTMC, F-75005 Paris, France}

\author{Julien Vidal}
\affiliation{Sorbonne Universit{\'e}, CNRS, Laboratoire de Physique Th{\'e}orique de la Mati{\`e}re Condens{\'e}e, LPTMC, F-75005 Paris, France}

%\date{\today}

\begin{abstract}

{We introduce low-energy effective models for dense configurations of vortices in the Kitaev honeycomb model. Specifically, we consider configurations of vortices in which vortex-free plaquettes form triangular lattices against a vortex-full background. Depending on the vortex density, these ``dual'' configurations belong to either one of two families classified by translation and inversion symmetry. As a function of a time-reversal symmetry breaking term, one family exhibits gapped phases with even Chern numbers separated by extended gapless phases, while the other exhibits gapped phases with even or odd Chern numbers, separated by critical points. We construct an effective model for each family, determine the parameters of these models by fitting the integrated density of states, and reproduce energy spectra and Chern numbers of the Kitaev honeycomb model. We also derive phase diagrams and determine these models' validity.}

\end{abstract}


\maketitle

\section{Introduction}

The characterization of free fermion phases via topological invariants represents a major achievement in the understanding of quantum matter~\cite{Schnyder08,Kitaev09,Ryu10,Chiu16}. Given the spatial dimension, these invariants may be determined through an analysis of global symmetries such as time-reversal (TRS), particle-hole, and chiral symmetries, in a classification scheme known as the tenfold way. In 2006, Kitaev introduced an exactly solvable spin-1/2 model on a honeycomb lattice that can be mapped to a free Majorana-fermion problem in a static $\mathbb{Z}_2$ gauge field~\cite{Kitaev06}. The gauge field results in the possibility of static $\pi$ fluxes, called vortices, in each  plaquette and the problem separates into independent vortex sectors. In the isotropic limit and in the vortex-free sector where the ground state lies (see below for details), the spectrum is gapless, but acquires a gap and a nonzero Chern number $\nu$ when TRS is broken~\cite{Kitaev06}. 
%However, explicitly introducing a magnetic field within the model gives rise to interactions between Majorana fermions. 
To break TRS while preserving the integrability of the model, Kitaev introduced a three-spin term that results in a next-nearest neighbor hopping of the Majorana fermions. The corresponding model, closely related to Haldane's model~\cite{Haldane88}, has $\nu=\pm 1$ and has been the subject of many studies (see, e.g., Refs.~\cite{Lahtinen08,Kamfor11,Lahtinen10,Lahtinen11,Kells11,Lahtinen12,Lahtinen14,Fu19,Zhang19,Zhang20,Fuchs20,Peri20,Grushin23,Cassella23}).

From the tenfold way, this model is a class-D topological superconductor in two dimensions, exhibiting an integer-valued Chern number $\nu\in\mathbb{Z}$ when the energy spectrum is gapped. Notably, Kitaev found that the bulk excitations of the model behave either as Abelian (even $\nu$) or non-Abelian (odd $\nu$) anyons, whose properties depend solely on $\nu\mod 16$~\cite{Kitaev06,Bernevig15}. These sixteen topological phases are suggested to occur in the fractional quantum Hall effect at filling factors with even denominator (such as the famous $5/2$)~\cite{Ma19}, and recently proposed honeycomb materials such as $\alpha$-Li$_2$IrO$_3$ and $\alpha$-RuCl$_3$ have been suggested to exhibit Kitaev-like quantum spin liquid phases~\cite{Trebst22}.

\begin{figure*}
    \centering
    \includegraphics[width=\textwidth]{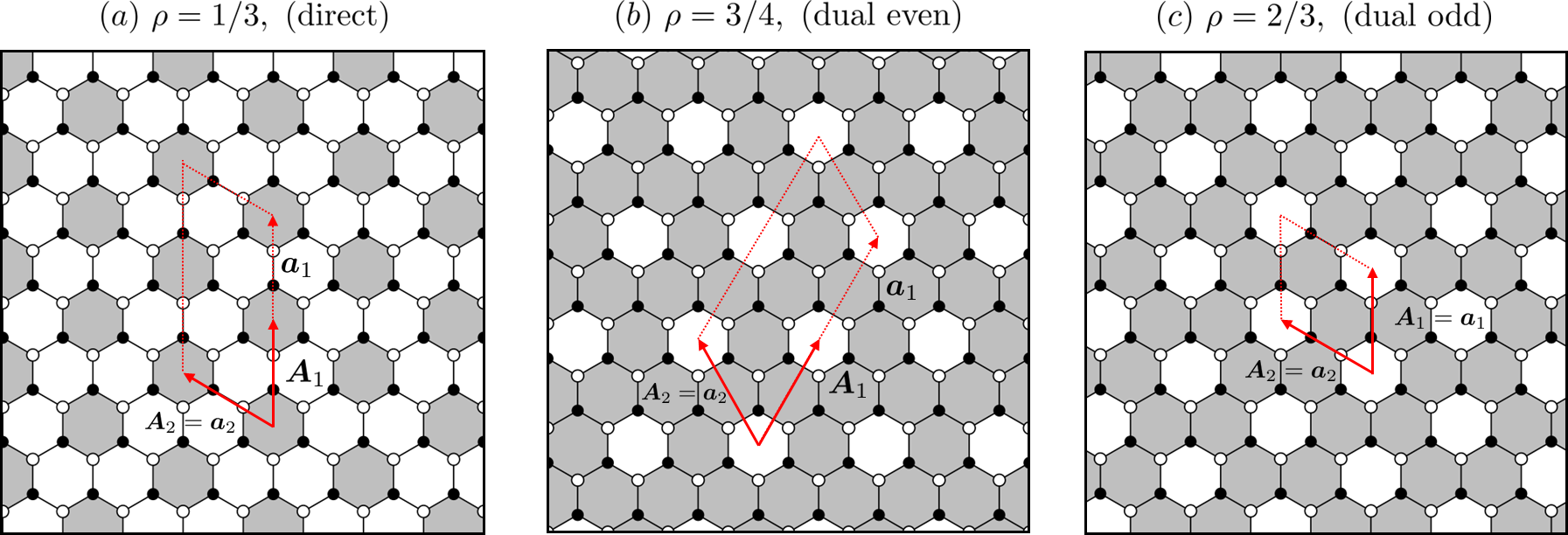}
    \caption{Examples of vortex arrangements with triangular symmetry: (a) direct, (b) dual even, and (c) dual odd.
    The vortex density is $\rho = 1/n$ for the direct configuration and $\rho = (n-1)/n$ for the dual configurations. Grey hexagons indicate plaquettes that host vortices ($w_{p} = -1$), while white hexagons indicate their absence ($w_{p} = +1$). Vectors ($\bm{A}_{1}$,$\bm{A}_{2}$) define the geometric unit cell (GUC), whereas ($\bm{a}_{1}$,$\bm{a}_{2}$) define the Hamiltonian unit cell (HUC) compatible with the periodicity of the $u_{jk}$'s.}
    \label{superlattices}
\end{figure*}

In the Kitaev model, the Chern number $\nu$ corresponds to the number of chiral Majorana edge modes propagating along the boundaries of the system and is associated with quantized thermal transport~\cite{Kitaev06}. A simple argument due to Volovik shows that if $\nu$ is odd, an isolated vortex traps an unpaired Majorana zero mode (MZM)~\cite{Volovik99}. In contrast, if $\nu$ is even, it traps a complex fermionic mode at finite energy, similar to the lowest Caroli-de Gennes-Matricon state in a $\nu=0$ superconductor~\cite{Caroli64}. This finite-energy mode can also be described as a pair of coupled MZMs.

Previous studies have analyzed the topological phases arising from triangular vortex configurations introduced on the honeycomb lattice~\cite{Lahtinen12,Lahtinen14,Zhang20,Fuchs20}. %While Kitaev originally assumed that this term is infinitesimally small and acts only in the vortex-free sector, we instead take this term to be valid in all vortex sectors and assume that it has arbitrary strength compared to the nearest neighbor interaction, as done in Ref.~\cite{Fuchs20}. 
 %The vortex configurations that we will focus on in this article can be separated into two distinct types: 
%In the first type, the  plaquettes of the honeycomb lattice that contain vortices are isolated from each other and form a triangular pattern against a background of vortex-free plaquettes. An example of this is illustrated in Fig.~\ref{superlattices}(a), where the plaquettes containing vortices are colored grey. 
These configurations are labeled by a vortex density $\rho=1/n$ with $n$ being an integer. Because of geometrical constraints, not every integer $n$ is accessible in a triangular lattice. For instance, $\rho=1/3$ is allowed [see Fig.~\ref{superlattices}(a)], but $\rho=1/5$ cannot be realized~\cite{Kamfor11,Fuchs20}. In the dilute limit, i.e., when $n\gg 1$, the spectra of these triangular vortex configurations only exhibit gapped phases with even Chern numbers, separated by gapless critical points as the three-spin coupling is varied. These numerical results are explained by an effective model that describes the hopping of Majorana fermions on a triangular lattice, whose sites correspond to the locations of the isolated vortices~\cite{Lahtinen12,Lahtinen14}. Indeed, in the vortex-free sector, the Chern number of the Kitaev honeycomb model is $\nu = \pm 1$, which implies that an isolated vortex traps an unpaired MZM~\cite{Kitaev06}. In the following, we will refer to these triangular vortex configurations as ``direct.''

The ``duals" to these triangular vortex configurations correspond to dense vortex configurations, with density $\rho=(n-1)/n$, where the plaquettes that do \textit{not} contain vortices form a triangular pattern against a background of plaquettes that all contain vortices. %Examples of this are shown in Figs.~\ref{superlattices}(b) and (c). 
These dense vortex configurations have recently been studied numerically in Refs.~\cite{Zhang20,Fuchs20}, and can be divided into two families according to their behavior with respect to translation and inversion symmetry. We shall refer to these two families as ``dual even'' (if $n$ is even such as $\rho=3/4, 11/12, 15/16,...$) and ``dual odd'' (if $n$ is odd such as $\rho=2/3,8/9,11/12,...$) in the following [see Figs.~\ref{superlattices}(b) and \ref{superlattices}(c)]~\cite{Kamfor11,Fuchs20}. The dual even family exhibits gapped phases with even Chern numbers, separated by extended gapless phases (spin metals with a Majorana-Fermi surface~\cite{Baskaran09,Tikhonov10}), whereas the dual odd family has gapped phases with even \textit{or odd} Chern numbers, separated by gapless points. These dual configurations are not describable by the effective model introduced in Refs.~\cite{Lahtinen12,Lahtinen14}. 

In this paper, we construct two different effective models describing the dual even and dual odd families. These models describe the hopping of pairs of coupled Majorana fermions on a triangular lattice, where the lattice sites correspond to the absences of vortices within a vortex sea. 
These effective models allow us to understand the variety of Chern numbers and the nature of gapless phases observed. 
The paper is organized as follows: in Sec.~\ref{Sec2} we briefly introduce the Kitaev honeycomb model in the Majorana fermion language and discuss the behaviors of the vortex-free and vortex-full sectors. In Sec.~\ref{Sec3}, we identify three distinct symmetry families for direct and dual vortex configurations based on translation and inversion symmetries.
In Sec.~\ref{Sec4}, we show that when the low-energy bands of the Kitaev honeycomb model are well-separated from the high-energy bands, they may be accurately described by an effective tight-binding model for all three of these families. We first review the construction of the low-energy model discussed in Refs.~\cite{Lahtinen12,Lahtinen14} for the direct case, and then introduce new effective models for the dual even and dual odd cases. To demonstrate the accuracy of these models, we directly compare their integrated density of states (IDOS) with that of the original Kitaev honeycomb model. %In addition, we derive phase diagrams for each symmetry family and discuss the validity of each effective model depending on the strength of $\kappa$ and the vortex density. 
Finally, in Sec.~\ref{Sec5}, we discuss our results, conclude and give perspectives. Appendices provide details on specific points such as the validity of the effective models (Appendix~\ref{app:1}) and some possible dispersion relations of vortex bands (Appendix~\ref{app:2}).

\section{Kitaev honeycomb model}
\label{Sec2}

\begin{figure}
    \centering
    \includegraphics[width=0.8\columnwidth]{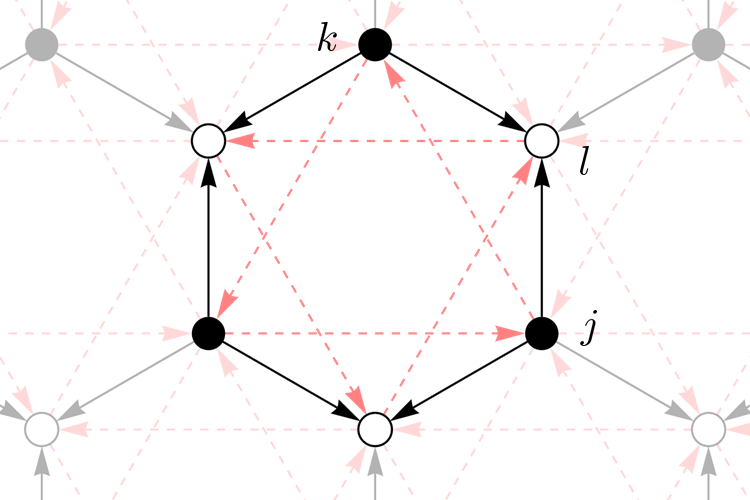}
    \caption{Standard gauge for the Hamiltonian defined in Eq.~\eqref{Ham}. Here a black arrow ($J$ hopping) pointing from site $k$ to site $l$ indicates that $u_{lk}=+1$, and consequently $u_{kl}=-1$. The two triangular sublattices of the honeycomb lattice are denoted by black and white colored sites. The dashed pink arrows ($\kappa$ hopping) represent the sign of the product $u_{jl}u_{lk}$ whenever the triplet $(k,l,j)$ is oriented clockwise: Whenever $(k,l,j)$ is a clockwise triplet of sites, and $u_{jl}u_{lk}=-1$, then the dashed pink arrow points from site $j$ to site $k$.}
    \label{honeycombschem}
\end{figure}

\begin{table*}
\begin{tabularx}{\textwidth}{ | Y || Y | Y || Y | Y|}
   
    \hline
%\thead{Background}
& $ \kappa = 0 $   &   $\kappa\ll J$
        &   $\kappa\gg J$  & $J = 0$                                              \\
   
   \hline
   \hline
\
Vortex-free ($\rho=0$)

Two bands
    &
    
    %Honeycomb lattice with $0$ flux
    
    $\Delta = 0$
    %
    %$\nu = \o$
    &

    $\Delta= 6\sqrt{3}\kappa$
    
    $\nu = +1$
     &  
        $\Delta= 2 J$
         
         $\nu = +1$
        &
        %Two decoupled triangular lattices  with alternating $\pi/2$ flux
        
        $\Delta=0$
       % 
        %$\nu = \o$
        \\
\hline
\ 
Vortex-full ($\rho = 1$)

Four bands
    &   
    %Honeycomb lattice with $\pi$ flux
    
    $\Delta = 0$
    %
    %$\nu = \o$
    &
    
    $\Delta\simeq 2\sqrt{3}\kappa$
    
    $\nu=+2$ 
        &   
        $\Delta\simeq 2\sqrt{3}\kappa$
    
        $\nu=-2$   
        &
        %Two decoupled triangular lattices with uniform $\pi/2$ flux
        
        $\Delta=2\sqrt{3}\kappa$
        
        $\nu=-2$                                                \\
   \hline
\end{tabularx}
\caption{Band gap $\Delta$ and Chern number $\nu$ (when $\Delta>0$) of the vortex-free and vortex-full sectors (backgrounds) of Eq.~\eqref{Ham} in the $\kappa\ll J$ and $\kappa\gg J$ limits.}
\label{tablebackgrounds}
    \end{table*}

We begin by recalling some essential facts about Kitaev's honeycomb model. The original spin-1/2 model can be reformulated in terms of an effective quadratic fermionic Hamiltonian given by~\cite{Kitaev06}
\begin{equation}
    H = \dfrac{i}{4}\sum_{j,k}A_{jk}c_{j}c_{k}.
    \label{Ham}
\end{equation}
Here $c_{i}$ are Majorana operators defined on the vertices of the honeycomb lattice such that $c_{i}^{\dagger} = c_{i}$ and \mbox{$\{c_{i},c_{j}\} = 2\delta_{ij}$}. The summation runs over all sites of the lattice, and $A$ is a real skew-symmetric matrix obeying the following rules:
\begin{equation}
    A_{jk} = \begin{cases}
    2J u_{jk}, & \text{if $j,k$ are nearest neighbors}
    \\ 2\kappa u_{jl}u_{lk}, & \text{if $(k,l,j)$ is oriented clockwise}
    \\ 0, & \text{otherwise}
    \end{cases}.
    \label{HamSkewSymmetric}
\end{equation}
Here the $u_{jk}$ parameters are $\mathbb{Z}_{2}$ gauge variables defined such that $u_{jk} = -u_{kj} = \pm 1$ for each link of the lattice. We assume here that the coupling $J$ is the same in all three directions, corresponding to the isotropic limit of the original spin model. This Hamiltonian describes noninteracting Majorana fermions coupled to a static $\mathbb{Z}_{2}$ gauge field. The term $J$ describes the nearest-neighbor hopping, whereas $\kappa$ denotes the strength of the three-spin term that breaks TRS, corresponding to next-nearest-neighbor hopping (see Fig.~\ref{honeycombschem}). In the following, without loss of generality, we will assume that $J$ and $\kappa$ are nonnegative. 
% Unless otherwise stated (i.e. a few occasions when $J=0$), energy units will be such that $J=1$.

% While it was originally assumed that the terms proportional to $\kappa$ arise as a leading order correction from an external magnetic field, and therefore $\kappa/J\ll 1$, we instead take the above model defined by Eqs.~\eqref{Ham} and \eqref{HamSkewSymmetric} as a valid starting point for analysis and allow for $\kappa$ to be arbitrarily strong compared to the nearest neighbor coupling. The nearest neighbor terms proportional to $J$ include only a single gauge variable $u_{jk}$, and therefore clearly satisfy the skew-symmetry condition $A_{jk} = -A_{kj}$. However, the next-nearest neighbor terms proportional to $\kappa$ include the product of two gauge variables $u_{jl}u_{lk}$ (where the site $l$ is connected to sites $j$ and $k$). As such, these terms require a specific orientation choice to ensure the skew-symmetry of the matrix $A$, and thus the hermiticity of the Hamiltonian. We choose to define $A_{jk} = +2\kappa u_{jl}u_{lk}$ when the triplet of site $(k,l,j)$ is oriented clockwise, i.e., the hopping from $k$ to $l$ and then from $l$ to $j$ travels in a clockwise direction around the  plaquette.

For each  plaquette $p$ of the honeycomb lattice, we define the $\mathbb{Z}_{2}$ variable \mbox{$w_{p} = \prod_{(j,k)\in p} u_{jk}={\rm e}^{i \phi_p}$}, where $\phi_p$ is the flux threading the plaquette. Here, for each plaquette, the pair of sites $(j,k)$ is taken such that $k$ always belongs to one sublattice of the honeycomb lattice, and $j$ belongs to the other. A value of $w_{p} = -1$ ($+1$) indicates a flux $\phi_p=\pi$ ($0$) in the plaquette $p$, meaning the presence (absence) of a vortex. Two sets of $\mathbb{Z}_{2}$ link variables $u_{jk}$ defined on the lattice are said to be equivalent if they correspond to the same configuration of plaquette variables, or ``vortex configuration." In analyzing Eq.~\eqref{Ham}, one can then restrict to a particular vortex sector labeled by the pattern of $w_{p}$'s. In Fig.~\ref{superlattices}, we show several examples of vortex configurations, where the $w_{p}=-1$ plaquettes are shaded grey. 
%For every vortex sector %consisting of a periodic arrangement of vortices, 
%the Hamiltonian of Eq.~\eqref{Ham} is quadratic, and thus is always readily diagonalizable via Bloch's theorem for a properly chosen unit cell.\\
%Two particular sectors will be important in the following: the vortex-free sector, with $w_{p} = +1$ for all plaquettes, and the vortex-full sector, in which $w_{p}=-1$ for all plaquettes of the honeycomb lattice. The first one has been studied in ~\cite{Kitaev06}, while the second one has been studied in ~\cite{Lahtinen10,Fuchs20}. 

Direct vortex configurations [see Fig.~\ref{superlattices} (a)] are defined by two vectors $\bm{A}_{1}$ and $\bm{A}_{2}$ that indicate the presence of a vortex at location $\bm{r} = m\, \bm{A}_{1} + n\, \bm{A}_{2}$ for $(m,n)\in\mathbb{Z}^2$. For dual configurations [see Figs.~\ref{superlattices}(b) and \ref{superlattices}(c)], we define the vectors $\bm{A}_{1}$ and $\bm{A}_{2}$ to indicate the absence of a vortex at position $\bm{r}$. In the following we shall refer to the parallelogram built from the vectors $\bm{A}_{1}$ and $\bm{A}_{2}$ as the geometric unit cell (GUC). It is not necessarily the same as the Hamiltonian unit cell (HUC), spanned by the Bravais lattice vectors $\bm{a}_{1}$ and $\bm{a}_{2}$, and which depends on the gauge choice for the $u_{jk}$ variables.
%Note that while $\bm{A}_{1}$ and $\bm{A}_{2}$ indicate the locations of vortices (or lack thereof), they are not necessarily the proper lattice vectors for the unit cell (HUC) of the Hamiltonian in Eq.~\eqref{Ham}. This is because the gauge choice for the $u_{jk}$ variables creating the corresponding vortex configuration may not be periodic with $\bm{A}_{1}$ and $\bm{A}_{2}$~\cite{Fuchs20}. 
%Finally, we will label different vortex configurations by the vortex density per unit area $\rho$, with $\rho=1/n$ for the direct configurations and $\rho=(n-1)/n$ for the dual configurations~\cite{Fuchs20}, where $n$ is an integer. 
%When $n$ is even (odd), we call this configuration dual even (dual odd).

We now discuss the physics of the vortex-free sector, for which $w_{p} = +1$ for all plaquettes ($\rho = 0$)~\cite{Kitaev06}, and of its dual, the vortex-full sector, where $w_{p}=-1$ for all plaquettes of the honeycomb lattice ($\rho = 1$)~\cite{Lahtinen10,Fuchs20}. They will constitute the backgrounds to our effective models. Their properties are summarized in Table~\ref{tablebackgrounds}.

\textit{Vortex-free sector. --} The standard gauge that realizes this configuration is represented in Fig.~\ref{honeycombschem}. The unit cell is made of one hexagonal plaquette (with zero flux) so that there are two bands. When $\kappa = 0$, the two bands touch at two Dirac points within the first Brillouin zone (same spectrum as graphene). Introducing a TRS-breaking term such that $\kappa\ll J$ opens a gap $\Delta=6\sqrt{3}\kappa$~\cite{Kitaev06,Haldane88}. The resulting Chern number is $\nu = 1$. As $\kappa$ increases, the system remains gapped, and in the $\kappa\gg J$ limit, $\Delta=2J$. As can be seen in Fig.~\ref{honeycombschem}, when $J = 0$ and $\kappa > 0$, the two triangular sublattices of the honeycomb are decoupled and each exhibit an alternating pattern of $\pm\pi/2$ flux per triangle. The resulting band structure is gapless along nodal lines.
%The energy spectrum is then simply the sum of the spectra of the two similar triangular lattices. 
%The spectrum is gapless, and features lines of nodes within the Brillouin zone. This implies that the system is in the ``spin-metal" phase of quantum spin liquids (QSLs).

% Because the Chern number of the vortex-free sector is odd, this implies that an isolated vortex plaquette of a non-vortex-free sector will bind a single localized unpaired MZM inside the bulk energy gap opened by $\kappa$. A consequence of this is that if we were to consider the case of a dilute vortex lattice against a vortex-free background, the resulting energy spectrum can be split into high-energy bands coming from the vortex-free background and low-energy bands coming from the tunneling of Majorana fermions between the zero modes. This can be seen, for example, in Fig.~\ref{IDOSschem}(a), where we plot the IDOS for an example of a triangular vortex configuration.

\begin{table*}
\begin{tabularx}{\textwidth}{ | Y || Y | Y | Y | Y |}
   
    \hline
\textbf{Translation sym.}
    &  $P_{w} = -1$   
    &   $P_{w} = -1$ 
    & $P_{w} = +1$
    & $P_{w} = +1$  \\
   %\hline
   \hline
Unit cells &  HUC $=2$ GUC
&   HUC $=2$ GUC
& HUC $=$ GUC 
& HUC $=$ GUC\\
\hline
Gapped phases with Chern number $\nu$ &  Even $\nu$ 
&   Even  $\nu$
& Even or odd $\nu$
& Even or odd $\nu$ \\
\hline \hline
\textbf{Inversion symmetry}
    &$\mathcal{P}^{2} = -\mathbbm{1}$    
    &$\mathcal{P}^{2} = +\mathbbm{1}$ 
    &$\mathcal{P}^{2} = -\mathbbm{1}$
    &$\mathcal{P}^{2} = +\mathbbm{1}$  \\
   %\hline
   \hline
Gapless phases &  Point-like
%($\mathcal{P}^{2} = -\mathbbm{1}$)
&   Extended
%($\mathcal{P}^{2} = -\mathbbm{1}$) 
& Point-like
%($\mathcal{P}^{2} = +\mathbbm{1}$)
& Extended\\

%($\mathcal{P}^{2} = +\mathbbm{1}$) \\
\hline \hline
Vortex configurations and density &  \textbf{Direct}

$\rho = 1/n$

(Vortex-full $n\to 1$)
&   \textbf{Dual even}

$\rho = (n-1)/n$, $n$ \textbf{even}
& \textbf{Dual odd}

$\rho = (n-1)/n$, $n$ \textbf{odd} 

(Vortex-free $n\to 1$)
&   
\

\large{?} 
\\
\hline
%Other lattices &    &    
%& 3-12~\cite{Yao07,Dusuel08}
%&   \\
%\hline
\end{tabularx}
\caption{Four families built from translation ($P_w=\pm 1$) and inversion ($\mathcal{P}^2=\pm \mathbbm{1}$) symmetries. The geometric (GUC) and Hamiltonian (HUC) unit cells  are indicated. The direct and dual configurations considered in this paper occupy three families. 
%Other lattices refer to the square-octagon (4-8) and triangle-dodecagon (3-12) lattices in different vortex sectors.
}
\label{tablefamilies}
    \end{table*}

%In this article we are chiefly concerned with the opposite situation, where isolated vortex-free plaquettes form dilute patterns against a vortex-full background. 
%The opposite situation, the vortex-full sector, in which $w_{p}=-1$ for all plaquettes of the honeycomb lattice, has been previously studied in Refs.~\cite{Lahtinen10,Fuchs20}. 
\textit{Vortex-full sector. --} The simplest gauge choice that achieves this configuration has a unit cell made of two hexagonal plaquettes (with $\pi$ flux each), which results in four bands. When $\kappa = 0$, the energy spectrum has four gapless Dirac cones~\cite{Lahtinen10}. Turning on a small $\kappa$ creates a gap $\Delta \simeq 2\sqrt{3}\kappa$ and $\nu = 2$. At $\kappa = J/2$, the gap closes. 
%with parabolic energy dispersion about two critical momenta. 
Continuing to increase $\kappa$ re-opens a gap and $\nu = -2$. In the $\kappa\gg J$ limit, the system remains gapped with $\Delta \simeq 2\sqrt{3} \kappa$. For $J = 0$ and $\kappa > 0$, the two triangular sublattices are disconnected. Each sublattice is similar to a Claro-Wannier model with uniform $\pi/2$ flux per triangle~\cite{Claro79}. This model has two bands of width $\sqrt{3}\kappa$ separated by a gap $\Delta=2\sqrt{3}\kappa$ and a Chern number $\nu=-1$.

\medskip

Throughout the rest of this paper, unless otherwise stated, we shall define our energy units such that $J=1$, with the assumption that $J$ is nonzero and positive.

\section{Translation and inversion symmetries}
\label{Sec3}
As shown in Ref.~\cite{Fuchs20}, direct and dual vortex configurations can be separated into three families based on translation and inversion symmetries. We review here these three families and their properties, which will motivate the introduction of the effective models.

Whether or not the gauge choice for the $u_{jk}$ variables is invariant under translations of $\bm{A}_{1}$ and $\bm{A}_{2}$ is found to depend on the parity of the vortex number $P_{w}$ inside the GUC~\cite{Fuchs20,Zhang20}. 
When $P_{w} = +1$, there is an even number of vortices in the GUC. In this case, there exists a gauge choice invariant under translations of $\bm{A}_{1}$ and $\bm{A}_{2}$ and the HUC is the same as the GUC~\cite{Kamfor11,Fuchs20}. By contrast, when $P_{w} = -1$, there is an odd number of vortices in the GUC, and the HUC must be double that of the GUC. Importantly, it has been shown that when $P_{w}=-1$, the system can only exhibit even-valued Chern numbers, while if $P_{w} = +1$ both even and odd Chern numbers are possible. The direct vortex configurations always exhibit $P_{w}=-1$, while the dual configurations can either exhibit $P_{w} = +1$ or \mbox{$P_{w} = -1$}~\cite{Fuchs20}. A vortex configuration with $P_{w} = +1$ is shown in Fig.~\ref{superlattices}(b), while vortex configurations with $P_{w}=-1$ can be seen in Figs.~\ref{superlattices}(a) and \ref{superlattices}(c). 

We can also analyze how the $u_{jk}$ gauge variables are affected under application of the inversion symmetry operator $\mathcal{P}= \mathcal{G} \mathcal{I}$, which is the product of a pure spatial inversion $\mathcal{I}$ and of a $\mathbb{Z}_2$ gauge transformation $\mathcal{G}$.
As discussed in Refs.~\cite{Fuchs20,Zhang20}, the square of this operator behaves as $\mathcal{P}^{2}=\pm\mathbbm{1}$. 
% If $\mathcal{P}^{2}=+\mathbbm{1}$, the gauge transformation is symmetric with $[\mathcal{I},\mathcal{G}]=0$ and $g_{-j}=g_{j}$. If instead $\mathcal{P}^{2}=-\mathbbm{1}$, the gauge transformation is antisymmetric with $\{\mathcal{I},\mathcal{G}\}=0$ and $g_{-j} = -g_{j}$. 
Extended gapless phases do not exist for vortex configurations with $\mathcal{P}^{2}=-\mathbbm{1}$.
Of the vortex configurations discussed so far, only the dual even exhibit $\mathcal{P}^{2} = +\mathbbm{1}$. The direct and the dual odd configurations, however, both exhibit $\mathcal{P}^{2} = -\mathbbm{1}$. 

In Table~\ref{tablefamilies}, we classify different families based on the signs of $P_{w} = \pm1$ and $\mathcal{P}^{2}=\pm\mathbbm{1}$. There are neither direct nor dual configurations for which $P_{w} = +1$ and $\mathcal{P}^{2}=+\mathbbm{1}$. 
%One immediate conclusion we may draw from Table~\ref{tablefamilies} is that while $P_{w} = \pm1$ and $\mathcal{P}^{2}=\pm\mathbbm{1}$ offer a total of four possible symmetry families, the triangular vortex configurations and dual configurations built from them only occupy three of these families. 
%In addition to the honeycomb, we also indicate other lattices (such as the square-octagon~\cite{Baskaran09} and triangle-dodecagon~\cite{Yao07,Dusuel08}) in which the Kitaev model was studied in specific vortex sectors. 
It was implicit in the analysis of Ref.~\cite{Fuchs20} that for the vortex configurations studied there, the gauge transformation can be chosen to have the periodicity of the Hamiltonian (i.e., that the operator $\mathcal{P}$ commutes with the translation operators defining the HUC). However, this might not be true for other vortex configurations on the honeycomb lattice or for the Kitaev model on other lattices, such as the square-octagon~\cite{Baskaran09} or the triangle-dodecagon\mbox{~\cite{Yao07,Dusuel08,Tikhonov10, Yao11}}.

\section{Effective low-energy models}
\label{Sec4}
\begin{figure*}
    \centering
    \includegraphics[width=\textwidth]{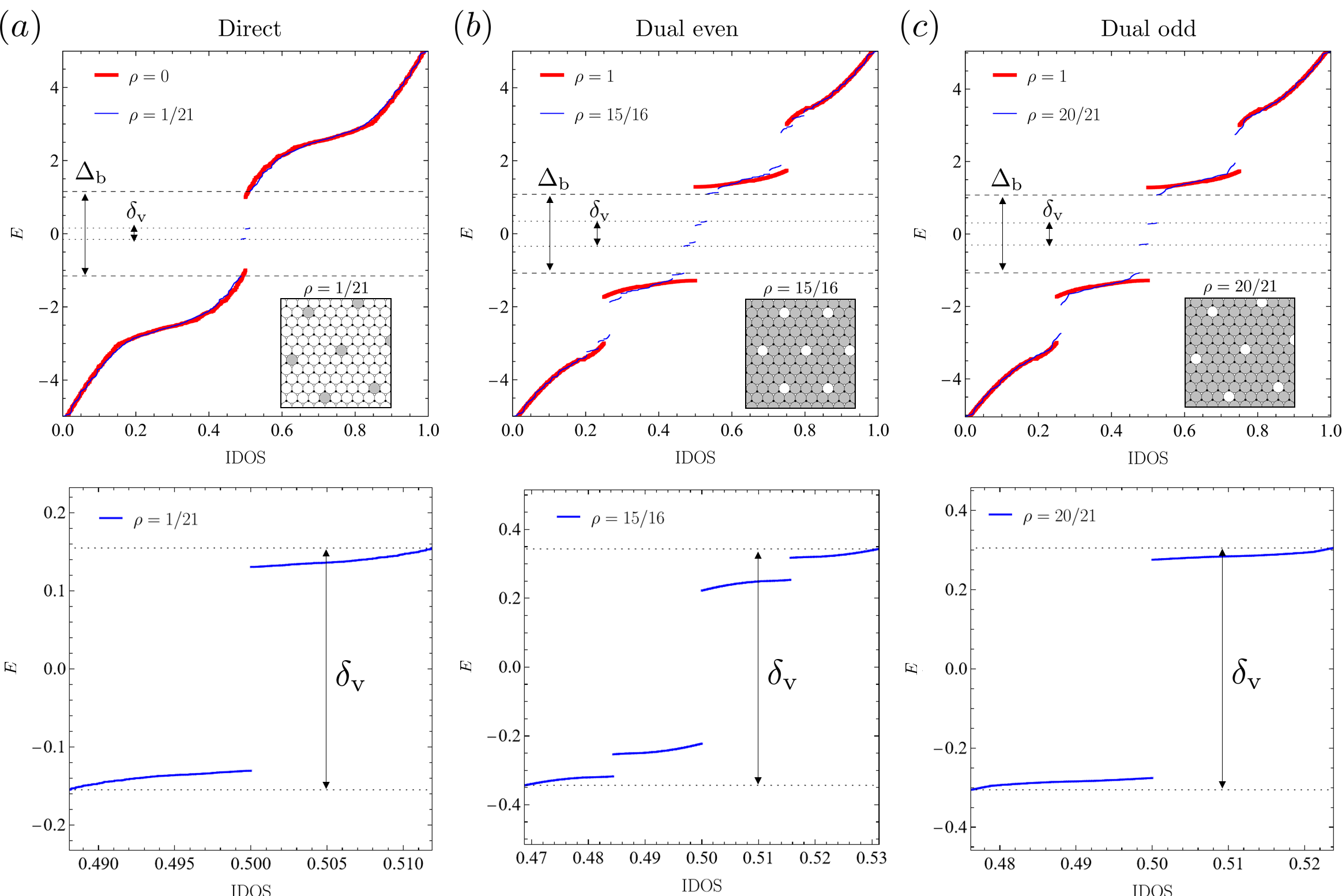}
    \caption{Energy vs IDOS of $H$ plotted in blue for: (a) direct ($\rho=1/21$), (b) dual even ($\rho=15/16$), and (c) dual odd ($\rho=20/21$), all with $\kappa = 1$ (energy units are such that $J=1$). In (a), the thick red line gives the IDOS of the vortex-free background, while for (b) and (c) it gives that of the vortex-full background. The dotted grey lines indicate the total bandwidth $\delta_{\rm v}$ of the vortex bands, while the dashed grey lines indicate the energy gap $\Delta_{\rm b}$ of the background bands. The second row shows a zoom on the vortex bands (there are two for the direct and dual odd configurations, and four for the dual even configurations).}
    \label{IDOSschem}
\end{figure*}
Previous works have shown that sufficiently sparse direct vortex configurations can be described by an effective Majorana tight-binding model that accurately reproduces their spectra and Chern numbers~\cite{Lahtinen12,Lahtinen14}. The intuition behind this model follows from analyzing the behavior of isolated vortices: given a vortex-free background, exchanging $u_{jk}\to -u_{jk}$ for a single link variable introduces two adjacent vortices into the system. When these vortices are sufficiently separated from each other, these $w_{p} = -1$ plaquettes each host a single unpaired MZM~\cite{Kitaev06}. The spectrum of a dilute direct vortex lattice against a vortex-free background can then be understood as arising from a set of high-energy bands coming from the background (``background bands") and a set of low-energy bands coming from the tunneling of Majorana fermions between these zero modes (``vortex bands"). This can be seen, for example, in Fig.~\ref{IDOSschem}(a) where we plot the IDOS of the Hamiltonian in Eq.~\eqref{Ham} exhibiting a direct vortex configuration with density $\rho = 1/21$. The IDOS $N(E)$ is defined as the number of states below a given energy $E$. Here, for practical use, we rather plot $E(N)$. The total Chern number of the system in this case can then be decomposed as $\nu = \nu_{\rm b} + \nu_{\rm v}$, where $\nu_{\rm b} = 1$ and $\nu_{\rm v}$ are the Chern numbers for the vortex-free background bands %(shown by dashed red lines) 
and the vortex bands, %(within the $\delta_{\rm v}$ bandwidth) 
respectively. 
%As demonstrated in Refs.~\cite{Lahtinen12,Lahtinen14}, including only the most relevant tunneling terms (i.e., nearest and next-nearest neighbor) in the effective tight-binding model predicts $\nu = 1 \pm 1$ and $\nu = 1 \pm 3$ when $\kappa$ and $J$ are positive, which is in agreement with the Chern numbers found within Eq.~\eqref{Ham} for triangular vortex configurations in the large-$n$ limit~\cite{Fuchs20}.

In this section, we expand upon the above analysis and derive low-energy effective models for the two families of dual vortex configurations. These cases are distinct from the direct configurations as an isolated $w_{p} = +1$ plaquette within a vortex-full background does not host an unpaired MZM. %In contrast, 
Instead, in the dual case, the even Chern number of the vortex-full background implies that an isolated ``dual vortex'' (i.e., a vortex-free plaquette surrounded by vortex-full plaquettes) will host a pair of coupled Majorana fermions equivalent to a single complex fermion mode with finite energy. The tunneling between these isolated finite-energy modes in the dilute limit then gives rise to a set of low-energy bands separated from the vortex-full background (we will still refer to these bands as ``vortex bands"). This can be seen in Figs.~\ref{IDOSschem}(b) and \ref{IDOSschem}(c), where we plot the IDOS for $\rho = 15/16$ (dual even) and $\rho = 20/21$ (dual odd). The number of low-energy bands (either two or four) depends on the symmetry family, as discussed below.

These effective models are valid as long as the total bandwidth $\delta_{\rm v}$ of the vortex bands is much smaller than the energy gap of the background bands $\Delta_{\rm b}$, as shown in Fig.~\ref{IDOSschem}. For justification and additional details, see Appendix~\ref{app:1}. 

\subsection{Warm-up: direct case
%vortex configurations
%: \\ $P_{w} = -1$, $\mathcal{P}^{2} = -\mathbbm{1}$
}
\begin{figure}
    \centering
    \includegraphics[scale=0.7]{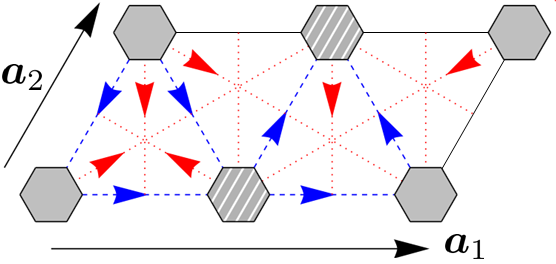}
    \caption{Direct.   Unit cell of the effective model defined in Eq.~\eqref{directtightbinding1}. Grey and hatched hexagons indicate the two inequivalent sites in the unit cell of the effective model. The directions of the blue dashed and red dotted arrows represent the sign of the gauge variables for the nearest ($t_1$) and next-nearest ($t_2$) neighbor hopping terms, respectively: if an arrow points from vortex site $k$ to vortex site $j$, then $s_{jk} = -s_{kj} = +1$.}
    \label{directmodel}
\end{figure}

We begin with a review of the construction of the low-energy model introduced by Lahtinen \textit{et al}. for the direct vortex configurations~\cite{Lahtinen12,Lahtinen14}. 
Defining the HUC vectors as $\bm{a}_{1} = 2\bm{A}_{1}$ and $\bm{a}_{2} = \bm{A}_{2}$,
we are then interested in providing an effective model describing Majorana fermions tunneling between vortices. The simplest corresponding Hamiltonian is given by
\begin{equation}
    H_{1} = it_{1}\sum_{\langle jk\rangle}s_{jk}\gamma_{j}\gamma_{k} + it_{2}\sum_{\langle\langle jk\rangle\rangle}s_{jk}\gamma_{j}\gamma_{k},
    \label{directtightbinding1}
\end{equation}
where $\gamma_i$ are Majorana operators defined on the $w_{p} = -1$ plaquettes of the honeycomb lattice and $(t_{1},t_{2})\in\mathbb{R}^2$ represent the hopping amplitudes between nearest and next-nearest-neighbor vortices, respectively. As explained in Ref.~\cite{Lahtinen12}, one needs these two terms to reproduce the Chern numbers observed. The $\mathbb{Z}_{2}$ gauge variables $s_{jk} = \pm 1$ are chosen such that the total flux accumulated by a Majorana fermion hopping counter-clockwise on a closed triangular circuit of $t_{1}$ links is $\phi_{1} = \pi/2$, the flux enclosed in a closed circuit of $t_{2}$ links is $\phi_{2} = -\pi/2$, and the flux enclosed in a closed circuit formed by both $t_{1}$ and $t_{2}$ links is $\phi_{1,2} = \pi/2$ (see Fig~\ref{directmodel}). In contrast with Ref.~\cite{Lahtinen12}, we choose a periodic pattern of $s_{jk}$ variables that both satisfies these requirements and gives rise to only two sites per unit cell. Utilizing the gauge choice defined in Fig.~\ref{directmodel}, we Fourier transform the Hamiltonian in Eq.~\eqref{directtightbinding1} to obtain

\begin{figure}
    \centering
    \includegraphics[scale=0.6]{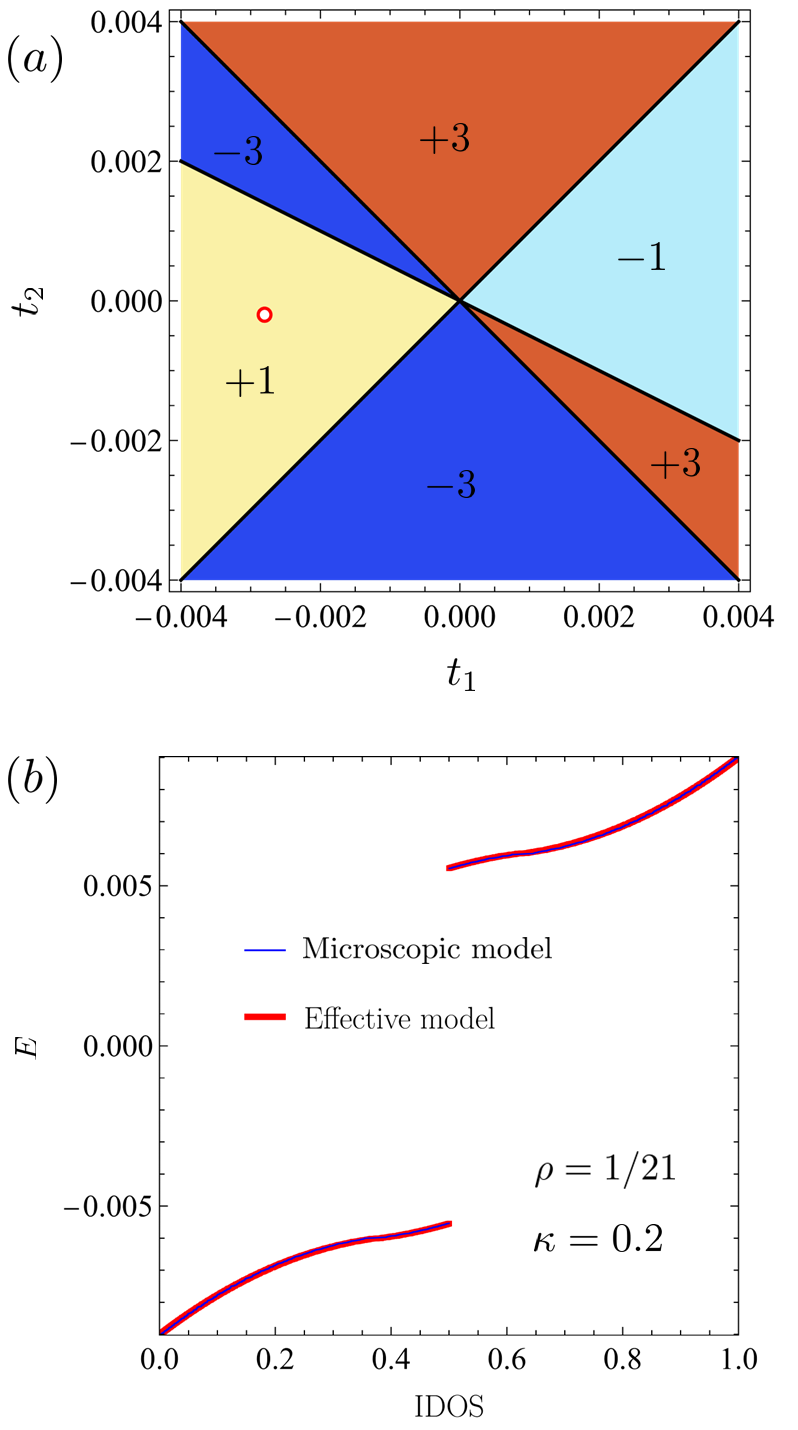}
    \caption{Direct.   (a) Phase diagram and Chern numbers $\nu_{\rm v}$ of the effective model given in Eq.~\eqref{directtightbinding1}, adapted from Ref.~\cite{Lahtinen12}. The phase boundaries are $t_{2}=t_{1}$, $-t_{1}$, and $-t_{1}/2$. (b) Comparison of the IDOS of the effective model and that of the \textit{vortex bands} of $H$ (both IDOS are normalized between 0 to 1). Here we have $\delta_{\rm v}/\Delta_{\rm b} \simeq 0.009$. The parameters are $t_{1} = -0.0028$, and $t_{2} = -0.0002$. These parameter values are indicated by the red dot in the phase diagram of (a), implying that $\nu_{\rm v} = +1$, in agreement with the microscopic model.}
    \label{directchernIDOS}
\end{figure}

\begin{equation}
    H_{1} = \sum_{\bm{k}} \Psi_{\bm{k}}^{\dagger}\mathcal{H}_{1}(\bm{k})\Psi_{\bm{k}}.
    \label{directtightbinding1f}
\end{equation}
Here $\Psi_{\bm{k}} =(\gamma_{\bm{k}},\overline{\gamma}_{\bm{k}})^{\rm T}$, such that $\gamma_{i}$ and $\overline{\gamma}_{i}$ represent the Majorana operators for the solid and striped vortices shown in Fig.~\ref{directmodel}, respectively, and
\begin{align}
    \begin{gathered}
        \mathcal{H}_{1}(\bm{k}) = \begin{pmatrix}
        f(\bm{k}) & g(\bm{k})
        \\ g(\bm{k})^{*} & -f(\bm{k})
        \end{pmatrix},
        \\ f(\bm{k}) = -2t_{1}\sin(\bm{k}\cdot\bm{a}_{2}) + 2t_{2}\sin(\bm{k}\cdot\bm{b}_{1}),
        \\ g(\bm{k}) = 2t_{1}[\sin(\bm{k}\cdot\bm{a}_{1}/2) - i\cos(\bm{k}\cdot\bm{a}_{3})]
        \\ \hspace{3cm} - 2t_{2}[\sin(\bm{k}\cdot\bm{b}_{2}) +i\cos(\bm{k}\cdot\bm{b}_{3})].
    \end{gathered}
    \label{fandg}
\end{align}
Here we have defined the nearest-neighbor directions as $\bm{a}_{1}/2$, $\bm{a}_{2}$, and $\bm{a}_{3} = \bm{a}_{2} - \bm{a}_{1}/2$, as well as the next-nearest neighbor directions as $\bm{b}_{1} = \bm{a}_{1} - \bm{a}_{2}$, $\bm{b}_{2} = \bm{a}_{2} + \bm{a}_{3}$, and $\bm{b}_{3} = \bm{a}_{2} + \bm{a}_{1}/2$. The above Majorana Hamiltonian in Eq.~\eqref{fandg} can be transformed into a Bogoliubov-de Gennes Hamiltonian by a $\bm{k}$-independent unitary transformation, revealing that the system is equivalent to an anisotropic $p_{x}+ip_{y}$ topological superconductor (see, e.g., Sec.~16.3 in Ref.~\cite{BernevigHughes}).

In Fig.~\ref{directchernIDOS}(a), we reproduce the phase diagram of this model from Ref.~\cite{Lahtinen12}, which features the possible vortex Chern numbers $\nu_{\rm v}=\pm1,\pm3$. We call $E_{\rm min}(t_{1},t_{2})$ and $E_{\rm max}(t_{1},t_{2})$ the minimum and maximum positive energy eigenvalues of Eq.~\eqref{fandg}, which are equal to
\begin{align}
    \begin{gathered}
        E_{\rm min}(t_{1},t_{2}) = \min_{\bm{k}}\sqrt{f(\bm{k})^{2} + |g(\bm{k})|^{2}},
        \\ E_{\rm max}(t_{1},t_{2}) = \max_{\bm{k}}\sqrt{f(\bm{k})^{2} + |g(\bm{k})|^{2}}.
    \end{gathered}
    \label{eminemax}
\end{align}
Here, we used these extrema to optimize the fitting parameters $t_1$ and $t_2$ by imposing the proper bound of the IDOS (they will also be used in the following section). To give an example of the accuracy of this model, in Fig.~\ref{directchernIDOS}(b) we plot the IDOS of the low-energy bands of the Kitaev honeycomb model in Eq.~\eqref{Ham} along with that of the effective model given in Eq.~\eqref{directtightbinding1f}. Here, we fit the parameters $t_{1}$ and $t_{2}$ by matching the IDOS (and Chern number) corresponding to Eq.~\eqref{directtightbinding1f} with that of Eq.~\eqref{Ham}. It is possible that multiple choices of $t_{1}$ and $t_{2}$ may result in the same IDOS as that of Eq.~\eqref{Ham}. In this case, as argued in Ref.~\cite{Lahtinen12}, the relative magnitudes of these parameters may be predicted by extracting $t_{1}$ and $t_{2}$ from the energy splitting of MZMs obtained by a numerical calculation with two spatially separated vortices in a vortex-free background~\cite{Lahtinen12}. While this method is less precise than exactly matching the IDOS, it has the advantage that it provides a unique solution. Therefore both approaches should be combined.

\subsection{Dual even case}
\begin{figure}
    \centering
    \includegraphics[scale=0.7]{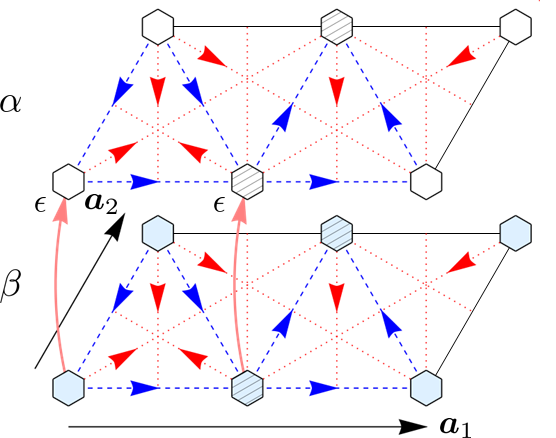}
    \caption{Dual even.  Unit cell of the effective model introduced in Eq.~\eqref{dualtightbinding2kspace}. Each ``dual vortex'' hosts two Majorana modes, which we interpret as two ``layers" that we color white and light blue. Here,  $\epsilon$ represents the on-site coupling of two modes on the same site, and the blue dashed and red dotted arrows represent the sign of the gauge variables for the nearest and next-nearest-neighbor hopping terms, respectively. The pattern of these gauge variables for both layers is the same as that given in Fig.~\ref{directmodel}.}
    \label{model34}
\end{figure}

In this section we introduce a low-energy model for dual even vortex configurations. 
The size of the HUC is double that of the GUC, and we set $\bm{a}_{1} = 2\bm{A}_{1}$, $\bm{a}_{2} = \bm{A}_{2}$, where the vectors $\bm{A}_{1}$ and $\bm{A}_{2}$ now generate the dual vortex lattice. Each of these isolated $w_{p} = +1$ plaquettes (i.e., a ``dual vortex'') hosts two Majorana modes at finite energy $\pm\epsilon$. 
%These modes are similar to the lowest-lying Caroli-de Gennes-Matricon states of a topological superconductor with even Chern number (including $\nu=0$)~\cite{Caroli64}. 
The splitting $2\epsilon$ has been calculated numerically and plotted in Fig.~6 of Ref.~\cite{Fuchs20}. It vanishes in the $J=0$ limit in which the two triangular sublattices of the honeycomb lattice are decoupled (see the discussion on the vortex-full sector in Sec.~\ref{Sec2}). This means that the two Majorana modes in the same ``dual vortex'' are uncoupled and therefore at zero energy. Inspired by this $J=0$ limit and by the effective model in the direct case [see Eq.~\eqref{directtightbinding1}], we construct a tight-binding model that features two ``layers" (or flavors) of Majorana modes. They tunnel between the dual vortices according to the following effective Hamiltonian:
\begin{eqnarray}
    H_{2} &=& it_{1}\sum_{\langle jk\rangle}[s_{jk}^{(\alpha)}\alpha_{j}\alpha_{k} + s_{jk}^{(\beta)}\beta_{j}\beta_{k}] \nonumber \\
    && + it_{2}\sum_{\langle\langle jk\rangle\rangle}[s_{jk}^{(\alpha)}\alpha_{j}\alpha_{k} + s_{jk}^{(\beta)}\beta_{j}\beta_{k}] + i\epsilon\sum_{j}\alpha_{j}\beta_{j}.\qquad
\label{dualtightbinding2}
\end{eqnarray}
Here,  $\alpha_{i}$ and $\beta_{i}$ are Majorana operators defined on what we refer to as the ``upper" and ``lower" layers, respectively. This is schematically shown in Fig.~\ref{model34}, where the upper layer is colored white and the lower layer is colored light blue. 

\begin{figure}
    \centering
    \includegraphics[scale=0.6]{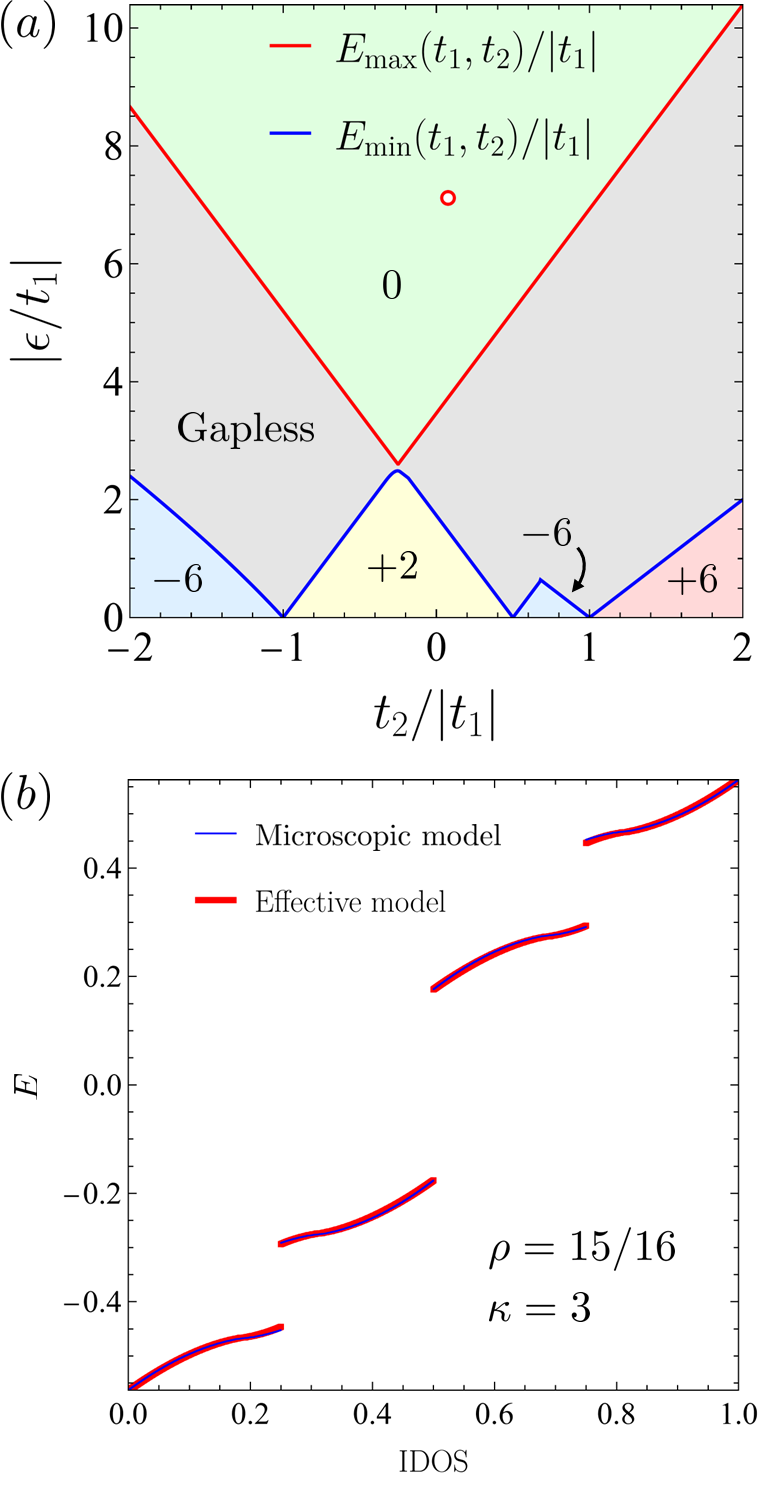}
    \caption{Dual even.   (a) Phase diagram and Chern numbers $\nu_{\rm v}$ of the effective model given in Eq.~\eqref{dualtightbinding2kspace}, where $t_{1} < 0$. $E_{\rm min}(t_{1},t_{2})$ and $E_{\rm max}(t_{1},t_{2})$ are defined in Eq.~\eqref{eminemax}. (b) Comparison of the IDOS of the effective model and that of the \textit{vortex bands} of $H$ (both IDOS are normalized between 0 to 1). Here, $\delta_{\rm v}/\Delta_{\rm b}\simeq 0.122$ and the parameters are $\epsilon = 0.370$, $t_{1} =  -0.052$, and $t_{2} = 0.004$. These parameter values are indicated by the red dot in the phase diagram of (a), implying that $\nu_{\rm v} = 0$, in agreement with the microscopic model.}
    \label{gaplesschern}
\end{figure}

In this construction, each isolated ``dual vortex'' in Fig.~\ref{superlattices}(b) hosts two Majorana modes: one belonging to $\alpha_j$, and one belonging to $\beta_j$. The on-site coupling between them is given by $\epsilon$, and at the current level we do not allow any interlayer tunneling between neighboring dual-vortex sites; nearest and next-nearest-neighbor tunneling occurs only between sites of the same layer. Including such terms would violate the $\mathcal{P}^{2} = +\mathbbm{1}$ symmetry of the microscopic model. As before, $s_{jk}^{(\alpha)} = \pm 1$ and $s_{jk}^{(\beta)} = \pm 1$ are $\mathbb{Z}_{2}$ gauge variables that determine the total enclosed flux of triangular circuits of dual vortices. From the $P_{w} = -1$ symmetry [which can be seen in Fig.~\ref{superlattices}(b)], the total flux enclosed when hopping counter-clockwise on a nearest-neighbor circuit of $t_{1}$ links is $\phi_1 = -\pi/2$, the flux enclosed in a next-nearest-neighbor circuit of $t_{2}$ links is $\phi_{2} = \pi/2$, and the flux enclosed in a circuit composed of both $t_{1}$ and $t_{2}$ links is $\phi_{1,2} = -\pi/2$. We may observe that this is simply the opposite of the flux pattern of the previous section. Noting that exchanging $\kappa \to -\kappa$ reverses the flux, we may adopt the same pattern of $s_{jk}$ variables defined in Fig.~\ref{directmodel}(b) for both layers of Eq.~\eqref{dualtightbinding2}. That is, we set $s_{jk}^{(\alpha)} = s_{jk}^{(\beta)}$. Doing so allows us to Fourier transform the Hamiltonian of Eq.~\eqref{dualtightbinding2} to find
\begin{equation}
    H_{2} = \sum_{\bm{k}} \Psi_{\bm{k}}^{\dagger}\begin{pmatrix}
    \mathcal{H}_{1}(\bm{k}) & i\epsilon\tau_{0}
    \\ -i\epsilon\tau_{0} & \mathcal{H}_{1}(\bm{k})
    \end{pmatrix}\Psi_{\bm{k}}.
    \label{dualtightbinding2kspace}
\end{equation}
Here,  $\Psi_{\bm{k}} = (\alpha_{\bm{k}},\overline{\alpha}_{\bm{k}},\beta_{\bm{k}},\overline{\beta}_{\bm{k}})^{\rm T}$, such that $\alpha_{\bm{k}}$ and $\overline{\alpha}_{\bm{k}}$ refer to the solid and striped white dual vortices, respectively (where solid and striped indicate the two inequivalent sites within the unit cell), $\beta_{\bm{k}}$ and $\overline{\beta}_{\bm{k}}$ refer to the solid and striped light blue dual vortices, respectively, $\tau_{0}$ is the identity matrix in the space of solid and striped dual vortices, and $\mathcal{H}_{1}(\bm{k})$ is defined in Eq.~\eqref{fandg}. As can clearly be seen, when $\epsilon = 0$ the four energy bands of Eq.~\eqref{dualtightbinding2kspace} consist of two copies of the bands of the previous model. Due to the lack of level repulsion resulting from the $\mathcal{P}^{2} = +\mathbbm{1}$ symmetry, these two pairs of bands begin to shift in opposite directions as $|\epsilon|$ is increased from zero. Whenever $E_{\rm min}(t_{1},t_{2}) < |\epsilon| < E_{\rm max}(t_{1},t_{2})$, two of these bands will overlap with each other at zero energy, leading to the appearance of a Majorana-Fermi surface in the Brillouin zone~\cite{Baskaran09,Tikhonov10}. When $|\epsilon|<E_{\rm min}(t_{1},t_{2})$, two copies of the valence band of Eq.~\eqref{fandg} will have negative energy, and the vortex Chern number of Eq.~\eqref{dualtightbinding2kspace} will simply be twice that of Eq.~\eqref{fandg}. When $|\epsilon|>E_{\rm max}(t_{1},t_{2})$, one valence band and one conduction band of the model in Eq.~\eqref{fandg} will be at negative energy. These bands each have opposite Chern numbers, and thus the total vortex Chern number of the system is zero. If we define $\nu^{(1)}$ as the vortex Chern number of the model defined in Eq.~\eqref{fandg}, the Chern number of the present model is thus given by
\begin{equation}
    \nu_{\rm v} = \begin{cases}
    2\nu^{(1)}, & \text{as $|\epsilon|<E_{\rm min}(t_{1},t_{2})$}
    \\ \text{gapless}, & \text{as $|\epsilon|\in[E_{\rm min}(t_{1},t_{2}),E_{\rm max}(t_{1},t_{2})]$}
    \\ 0, & \text{as $|\epsilon|>E_{\rm max}(t_{1},t_{2})$}
    \end{cases}.
    \label{nu1cases}
\end{equation}

In Fig.~\ref{gaplesschern}(a), we plot the phase diagram of Eq.~\eqref{dualtightbinding2kspace} for $t_1<0$ ($t_{1} > 0$ simply changes the sign of the Chern numbers and mirrors the plot). For all values of $t_{1}$ and $t_{2}$ we find that we always have $E_{\rm min}(t_{1},t_{2}) < E_{\rm max}(t_{1},t_{2})$ [see Eq.~(\ref{eminemax})], i.e., there is always a Majorana-Fermi surface for certain values of $\epsilon$. Because the vortex-full background has the possible Chern numbers $\nu_{\rm b} = \pm2$, the above model thus predicts that the total Chern number can take the values $\nu = 0,\pm2,\pm4$, and $\pm 8$. The values $\nu = 0,\pm 2,$ and $\pm 4$ have indeed been observed, but not $\pm 8$~\cite{Fuchs20}.

In Fig.~\ref{gaplesschern}(b), we plot the IDOS of the low-energy bands of $H$ along with that of the effective Hamiltonian $H_2$. As before, we fit the parameters $\epsilon$, $t_{1}$, and $t_{2}$  by matching the IDOS and Chern number with Eq.~\eqref{Ham}. %If the Chern number is not known beforehand, it is possible to evaluate the three parameters via numerically analyzing a vortex configuration consisting of three dual-vortices in a vortex-full sea. {\red This is a bit short.}

\subsection{Dual odd case}
Next we introduce a low-energy model for the dual odd configurations. According to Table~\ref{tablefamilies}, in this case the $P_{w} = +1$ symmetry implies that there is an even number of vortices within each GUC, as can be seen in the example shown in Fig.~\ref{superlattices}(c) for $\rho = 2/3$. Unlike the previous two families, for these configurations the HUC and GUC are equivalent to one another, and we therefore set \mbox{$\bm{a}_{1} = \bm{A}_{1}$} and \mbox{$\bm{a}_{2} = \bm{A}_{2}$}.

We wish to construct a tight-binding model that features two layers of Majorana modes that tunnel between dual vortices located. However, in anticipation of the $\mathcal{P}^{2} = -\mathbbm{1}$ symmetry constraint, this time we shall allow interlayer tunneling between neighboring dual-vortex sites. The Hamiltonian is then
\begin{eqnarray}
    H_{3} =&& it_{1}\sum_{\langle jk\rangle}[s_{jk}^{(\alpha)}\alpha_{j}\alpha_{k} + s_{jk}^{(\beta)}\beta_{j}\beta_{k}]
    \nonumber \\ 
    &+& it_{2}\sum_{\langle\langle jk\rangle\rangle}[s_{jk}^{(\alpha)}\alpha_{j}\alpha_{k} + s_{jk}^{(\beta)}\beta_{j}\beta_{k}] 
     \nonumber \\ 
    &+& it_{\perp 1}\sum_{\langle jk\rangle}[s_{jk}^{(\perp)}\alpha_{j}\beta_{k} + \widetilde{s}_{jk}^{(\perp)}\beta_{j}\alpha_{k}]
     \nonumber \\ 
    &+& it_{\perp 2}\sum_{\langle\langle jk\rangle\rangle}[s_{jk}^{(\perp)}\alpha_{j}\beta_{k} + \widetilde{s}_{jk}^{(\perp)}\beta_{j}\alpha_{k}]  \nonumber \\
    &+& i\epsilon\sum_{j}\alpha_{j}\beta_{j}.
\label{dualtightbinding3}
\end{eqnarray}
Here,  $(t_{1},t_{2})\in\mathbb{R}^2$ represent the hopping amplitudes of modes between nearest and next-nearest-neighbor sites in the same layer, respectively. In contrast, the new terms $(t_{\perp 1},t_{\perp 2})\in\mathbb{R}^2$ represent the interlayer hopping amplitudes between nearest and next-nearest neighbors, respectively. This Hamiltonian then has the four sets of gauge variables: $s^{(\alpha)}_{jk}$, $s^{(\beta)}_{jk}$, $s^{(\perp)}_{jk}$, and $\widetilde{s}^{(\perp)}_{jk}$. To obey the $\mathcal{P}^{2} = -\mathbbm{1}$ symmetry we must set $s^{(\beta)}_{jk} = -s^{(\alpha)}_{jk}$, which leaves us with three available sets of gauge variables to choose in order to analyze the model, along with the signs of the hopping parameters and $\epsilon$.
\begin{figure}
    \centering
    \includegraphics[scale=0.7]{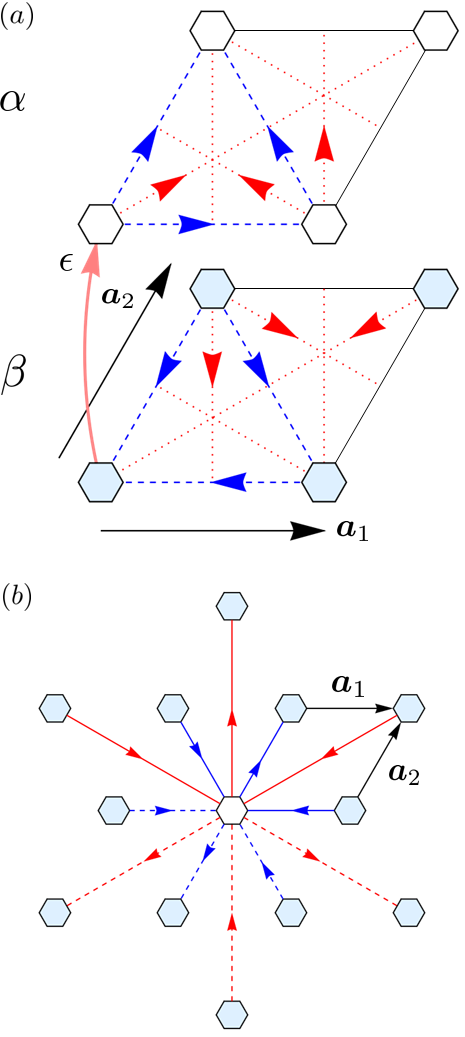}
    \caption{Dual odd.  Unit cell of the effective model defined in Eq.~\eqref{dualtightbinding3kspace}. (a) The blue dashed and red dotted arrows represent the sign of the gauge variables for the nearest $(t_1)$ and next-nearest-neighbor $(t_2)$ \textit{intralayer} hopping terms, respectively. The upper layer corresponds to the gauge variables $s^{(\alpha)}_{jk}$, while the lower layer corresponds to the gauge variables $s^{(\beta)}_{jk} = -s^{(\alpha)}_{jk}$. (b) Schematic of the \textit{interlayer} hopping amplitudes in Eq.~\eqref{dualtightbinding3kspace} (top view). The terms $t_{\perp 1}$ and $t_{\perp 2}$ are colored blue and red, respectively. Solid arrows refer to the choice of signs for the $\widetilde{s}^{(\perp)}_{jk}$ gauge variables, while dashed arrows refer to $s^{(\perp)}_{jk}$.}
    \label{model23}
\end{figure} 
\begin{figure}
    \centering
    \includegraphics[scale=0.6]{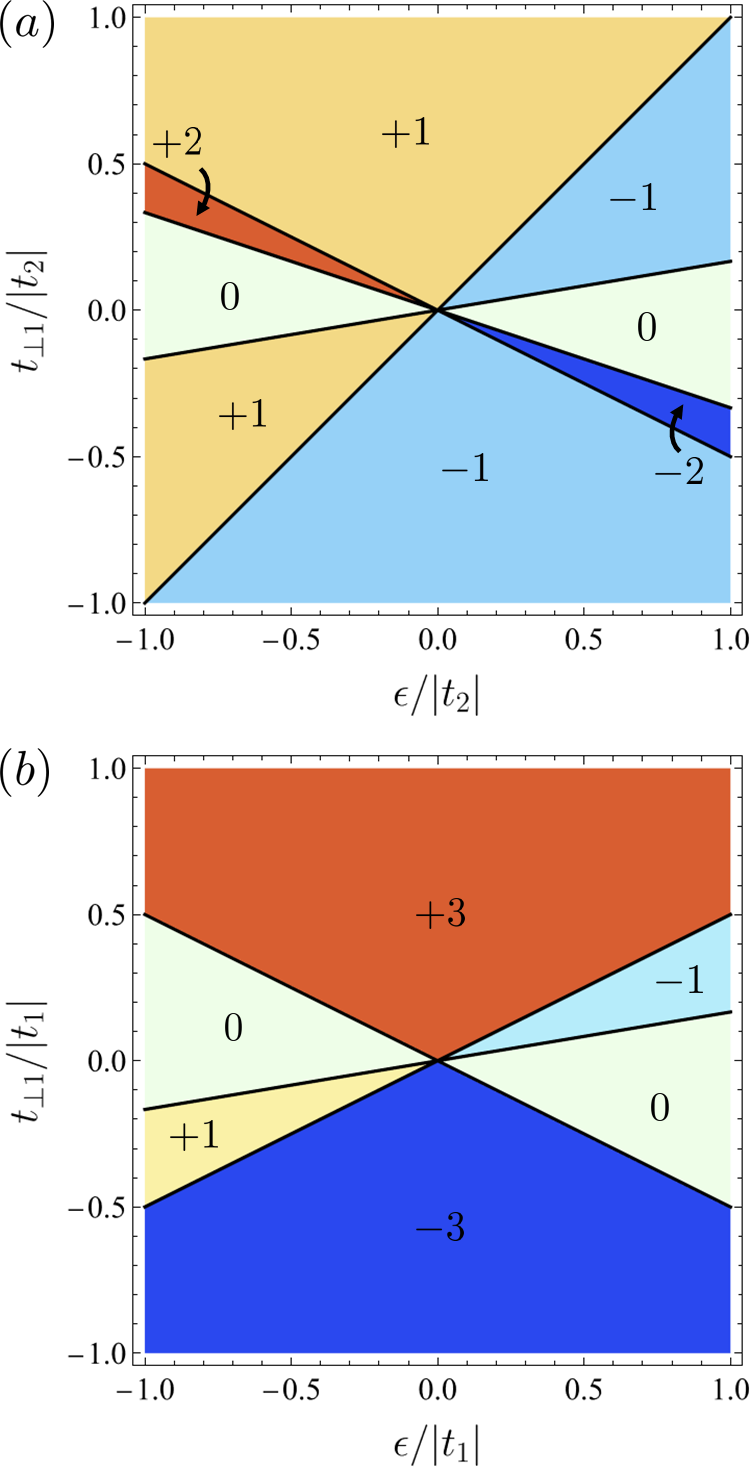}
    \caption{Dual odd.  Phase diagram and Chern numbers $\nu_{\rm v}$ of the model given in Eq.~\eqref{dualtightbinding3kspace}. (a) $t_{1} = 0$ and $t_{2} = t_{\perp 2}$. The phase boundaries are $t_{\perp 1} = \epsilon/6, \epsilon, -\epsilon/2$, and $-\epsilon/3$. (b) $t_{2} = 0$ and $t_{1} = t_{\perp 2}$. The phase boundaries are $t_{\perp 1} = \epsilon/6$ and $\pm \epsilon/2$.}
    \label{zeropichern}
\end{figure}

The original honeycomb model in Eq.~\eqref{Ham} shows that for this family the accumulated flux in a triangular circuit of nearest-neighbor dual vortices is found to be always either $\phi = 0$ or $\phi = \pi$. Unlike in the previous two sections, the gauge variables of Eq.~\eqref{dualtightbinding3} cannot be chosen in such a way that recreates this pattern. Indeed, for Majorana fermions, triangular circuits always carry a flux of $\pm\pi/2$. However, as schematically shown in Fig.~\ref{model23}, the size of the HUC implies that all gauge choices of the $s^{(\alpha)}_{jk}$, $s^{(\perp)}_{jk}$, and $\widetilde{s}^{(\perp)}_{jk}$ variables result in patterns of alternating $\pm\pi/2$ flux within the nearest-neighbor triangular circuits, which is the closest a model of Majorana fermions can be to obtaining a flux of $0$ or $\pi$. As such, we are therefore unrestricted in our choice of gauge variables. Taking the gauge choices for $s^{(\alpha)}_{jk}$, $s^{(\perp)}_{jk}$, and $\widetilde{s}^{(\perp)}_{jk}$ defined in Fig.~\ref{model23}, we may Fourier transform the Hamiltonian to find
\begin{equation}
    H_{3} = \sum_{\bm{k}}\Psi_{\bm{k}}^{\dagger}\mathcal{H}_{3}(\bm{k})\Psi_{\bm{k}},
    \label{dualtightbinding3kspace}
\end{equation}
where here, $\Psi_{\bm{k}} = (\alpha_{\bm{k}},\beta_{\bm{k}})^{\rm T}$, and
\begin{align}
    \begin{gathered}
        \mathcal{H}_{3}(\bm{k}) = \begin{pmatrix}
        \widetilde{f}(\bm{k}) & \widetilde{g}(\bm{k})
        \\ \widetilde{g}(\bm{k})^{*} & -\widetilde{f}(\bm{k})
        \end{pmatrix},
        \\ \widetilde{f}(\bm{k}) = 2t_{1}\sum_{j=1}^{3}\sin(\bm{k}\cdot \bm{a}_{j}) + 2t_{2}\sum_{j=1}^{3}\sin(\bm{k}\cdot \bm{b}_{j}),
        \\ \widetilde{g}(\bm{k}) = i\epsilon + i2t_{\perp 1}[\cos(\bm{k}\cdot\bm{a}_{1}) - \cos(\bm{k}\cdot\bm{a}_{2}) + \cos(\bm{k}\cdot\bm{a}_{3})]
        \\ \hspace{1cm} - 2t_{\perp 2}[\sin(\bm{k}\cdot\bm{b}_{1}) - \sin(\bm{k}\cdot\bm{b}_{2}) + \sin(\bm{k}\cdot\bm{b}_{3})].
    \end{gathered}
    \label{ftandgt}
\end{align}
The nearest-neighbor directions are defined as $\bm{a}_{1}$, $\bm{a}_{2}$, and $\bm{a}_{3} = \bm{a}_{2} - \bm{a}_{1}$, while the next-nearest-neighbor directions are defined as $\bm{b}_{1} = \bm{a}_{1} + \bm{a}_{2}$, $\bm{b}_{2} = \bm{a}_{2} + \bm{a}_{3}$, and $\bm{b}_{3} = \bm{a}_{3} - \bm{a}_{1}$. The above Majorana Hamiltonian in Eq.~\eqref{ftandgt} can be transformed into a Bogoliubov-de Gennes Hamiltonian by a $\bm{k}$-independent unitary transformation, showing that the model is close to a $p_x+ip_y$ superconductor on a triangular lattice. This could have been anticipated given that a ``dual vortex'' hosts a single complex fermionic mode, that the energy spectrum must have particle-hole symmetry, and that the gapped bands carry nonzero Chern numbers.
\begin{figure}
    \centering
    \includegraphics[scale=1]{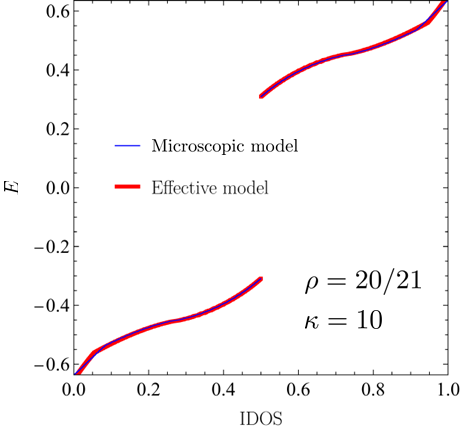}
    \caption{Dual odd.  Comparison of the IDOS of the effective model given in Eq.~\eqref{dualtightbinding3kspace} and that of the \textit{vortex bands} of $H$ (both IDOS are normalized between 0 to 1). Here,  \mbox{$\delta_{\rm v}/\Delta_{\rm b} \simeq 0.038$} and the parameters are $\epsilon = 0.395$, $t_{1} = 0.095$, $t_{\perp 1} = -0.043$, and all the other terms zero (note that these values do not correspond to one of the phase diagrams given in Fig.~\ref{zeropichern}). For an effective model given by these parameters we calculate the vortex Chern number to be $\nu_{\rm v} = 0$, in agreement with the microscopic model.}
    \label{zeropiIDOS}
\end{figure}

In Figs.~\ref{zeropichern}(a) and \ref{zeropichern}(b), we plot the phase diagrams of this model as $t_{1} = 0$, $t_{2} = t_{\perp 2}$ and $t_{2} = 0$, $t_{1} = t_{\perp 2}$, respectively, which collectively exhibit the Chern numbers $\nu_{\rm v} = 0,\pm1, \pm 2$, and $\pm 3$. Thus, knowing that the vortex-full background exhibits $\nu_{\rm b} = \pm 2$, the total Chern numbers arising from the above effective model are $\nu = 0, \pm1, \pm 2, \pm 3, \pm 4$ and $\pm 5$. This is in agreement with what is observed in Ref.~\cite{Fuchs20}, which, for this family, showcases all of these Chern numbers except $\nu =\pm 4$. We also note that the above model is just one of several possible choices for the gauge variables $s^{(\alpha)}_{jk}$, $s^{(\perp)}_{jk}$, and $\widetilde{s}^{(\perp)}_{jk}$, and other choices may give rise to additional Chern numbers. In total, the model in Eq.~\eqref{ftandgt} is determined by the five parameters $\epsilon, t_{1}, t_{2}, t_{\perp 1}$, and $t_{\perp 2}$, along with the choice of four signs. These are the signs in front of $\cos(\bm{k}\cdot\bm{a}_{2})$, $\cos(\bm{k}\cdot\bm{a}_{3})$, $\sin(\bm{k}\cdot\bm{b}_{2})$ and $\sin(\bm{k}\cdot\bm{b}_{3})$ in Eq.~\eqref{ftandgt}. All other signs can be absorbed into the five parameters.

In Fig.~\ref{zeropiIDOS}, we plot the IDOS of the low-energy bands of the Kitaev honeycomb model in Eq.~\eqref{Ham} along with that of the effective model in Eq.~\eqref{dualtightbinding3kspace}. The five parameters $\epsilon$, $t_{1}$, $t_{2}$, $t_{\perp 1}$, and $t_{\perp 2}$ are found by assuming a relevant gauge and matching the IDOS and Chern number with Eq.~\eqref{Ham}. If the Chern number of the model is not known beforehand, then multiple sets of these parameters may give rise to the same IDOS. In this case, unlike the previous two symmetry families, there is no clear way to predict the relative magnitudes of these five parameters by studying a two ``dual vortex'' problem numerically.

In Appendix~\ref{app:2}, we show that not only the IDOS but also the dispersion relation is well-reproduced by this effective model. As particularly challenging cases, we consider gapless critical points separating gapped phases with different Chern numbers.

\section{Discussion and perspectives}
\label{Sec5}
In this paper, we extended the results of previous works by introducing low-energy models that can describe both sparse triangular (direct) and dual vortex configurations in Kitaev's honeycomb model. In the direct case, featuring a triangular pattern of vortices against a vortex-free background, the behavior of the Kitaev honeycomb model may be interpreted as arising from a combination of the background and of the tunneling of single MZMs between vortex sites. In the dual case, a triangular pattern of vortex-absences is instead superimposed on top of a vortex-full background. These isolated ``dual vortices" each contain a pair of coupled Majorana modes, and we have shown that the properties of the Kitaev honeycomb model may be interpreted as arising from a combination of the vortex-full background and of two ``layers" of finite-energy Majorana modes that tunnel to other dual-vortex sites in either an intralayer or interlayer fashion. 

In both the direct and dual cases, the phase diagram of the Kitaev honeycomb model is controlled by the TRS breaking term $\kappa$ that we allow to be arbitrarily strong compared to the nearest-neighbor interaction $J$. For certain values of $\kappa/J$, the low-lying energy bands of the microscopic model are well-separated from the high-energy bands, and in these cases we have demonstrated that the low-energy bands may be accurately modeled with our effective descriptions. In addition, the high-energy bands in the microscopic model resemble the spectra of the vortex-free or vortex-full backgrounds. Furthermore, for both the direct and dual configurations, we have found that the total Chern number of the microscopic model $\nu$ is equal to the sum of the background bands' Chern number $\nu_{\rm b}$ and the low-energy bands' Chern number $\nu_{\rm v}$, i.e., $\nu = \nu_{\rm b} + \nu_{\rm v}$. The Chern numbers predicted by our effective models are in very good agreement with what has been observed in Ref.~\cite{Fuchs20}. 
%the Kitaev honeycomb model of Eq.~\eqref{Ham} for all configurations with densities $\rho = 1/n$ and $\rho = (n-1)/n$ such that $n\leq 28$ and $0\leq \kappa/J\leq 10$, as well as the case of $J = 0$ and $\kappa\neq 0$~\cite{Fuchs20}.

We note that while the parameters of our effective models are found by directly fitting the IDOS against that of the Kitaev honeycomb model, it would be worthwhile to develop a deeper connection between the parameters of each model. The dual odd family, for instance, has a large number of degrees of freedom with five parameters and four signs of gauge variables to choose from. Knowing how these parameters depend on the TRS breaking term $\kappa$, for instance, may help us in better understanding the symmetries and topological properties of the Kitaev honeycomb model.

Among perspectives, one could think of using the above effective models to study, in detail,  the anyonic excitations predicted by the sixteenfold way~\cite{Kitaev06}. Preliminary steps in this direction have been taken in Ref.~\cite{Zhang20}. One could test the fusion and braiding properties of these excitations similarly to what has been done in the vortex-free sector in Ref.~\cite{Lahtinen11}. 

In addition, there are several open questions raised by this work that should be addressed. For example, is it possible to find a phase with Chern number $\nu = \pm 7$ mod 16 by considering nontriangular vortex configurations~\cite{Fuchs20}? Are there models that realize the fourth family (see Table~\ref{tablefamilies}) of inversion-symmetric and periodic lattices that is expected to feature extended gapless phases and even or odd Chern numbers, or is there a general argument showing that this is impossible? What happens in cases where inversion symmetry does not commute with the translation operators defining the HUC? More generally, can we use other crystalline symmetries than inversion to classify all type of behavior (parity of Chern number, nature of gapless phases) in the Kitaev model in different vortex sectors and on different lattices (the honeycomb but also the square-octagon~\cite{Baskaran09} or the triangle-dodecagon~\cite{Yao07,Dusuel08,Tikhonov10, Yao11} or the pentaheptite lattice~\cite{Peri20}, for example)? 

\acknowledgements
 We thank Surajit Basak for many fruitful discussions. D. J. A. acknowledges LPS Orsay, where the writing of this manuscript was completed, for hospitality. 
 We acknowledge the financial support of  the Quantum Information Center at Sorbonne Universit\'e (QICS) via an Emergence SU grant for the TQCNAA project and also that of the Domaine d'Int\'er\^et Majeur (DIM) QuanTip of the \^{I}le-de-France region.

\appendix

\section{Validity of the effective models}
\label{app:1}

In this Appendix we discuss the validity of our effective models. For a given vortex configuration, either direct or dual, we may define $\delta_{\rm v}$ as the total bandwidth of the low-energy vortex bands: two bands for direct or dual odd, four bands for dual even. We may then define $\Delta_{\rm b}$ as the energy gap of the next-highest bands. These parameters are defined explicitly in Fig.~\ref{IDOSschem}, where the magnitudes of $\delta_{\rm v}$ and $\Delta_{\rm b}$ are indicated by the dotted and dashed lines, respectively.

At small $\kappa$, the energy splitting of two isolated MZMs separated by a distance $d$ is qualitatively given by $t(d,\kappa)\simeq \Delta_{\rm b}e^{-d/\xi}\cos(k_{\rm F}d+\gamma)$, where $d=|\bm{A}_{1}| = |\bm{A}_{2}|$ is the vortex separation, $\xi$ is the superconducting coherence length ($\xi\sim W_{\rm b}/\Delta_{\rm b}$ with $W_{\rm b}$ the total bandwidth of the background), $k_F$ is the distance from the center of the Brillouin zone to the Dirac points and $\gamma$ is a phaseshift~\cite{Cheng09,Lahtinen12}. These Friedel-like oscillations are a hallmark of the two-vortex solution of isolated MZMs, and naturally appear in other systems exhibiting Ising anyons, such as the Moore-Read quantum Hall state, $p_{x} + ip_{y}$ superconductors, and topological nanowires. The oscillations are related to the change in fermionic parity and to the fact that MZMs behave as Ising anyons $\sigma$ that have the possibility to fuse either in the vacuum $1$ or in a fermion $\psi$ according to the well-known fusion rule $\sigma \times \sigma = 1 + \psi$ (see Fig.~1 in Ref.~\cite{Lahtinen12}). We may estimate the total vortex bandwidth $\delta_{\rm v}$ for the direct vortex configurations by $\delta_{\rm v}\sim |t(d,\kappa)|\sim \Delta_{\rm b}e^{-d/\xi}$.  Dilute vortex configurations have MZMs well-separated from one another, i.e., $d\gg\xi$. This implies that $\delta_{\rm v}\ll\Delta_{\rm b}$, which we take as a necessary condition for the validity of the effective model for the direct configurations.

\begin{figure}
    \centering
    \includegraphics[scale=0.6]{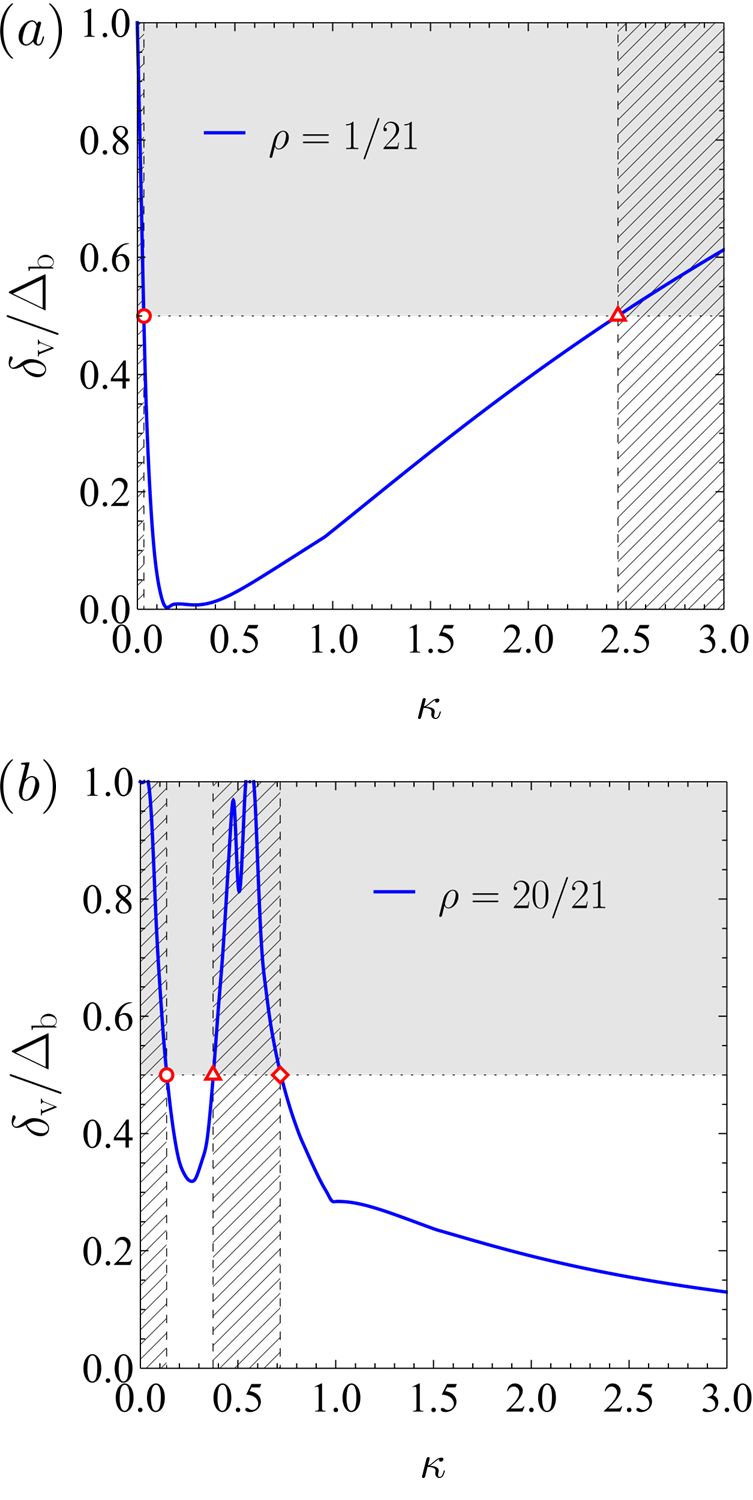}
    \caption{Ratio of the low-energy bandwidth $\delta_{\rm v}$ and the energy gap $\Delta_{\rm b}$ as  a function of $\kappa$ for a direct ($\rho = 1/21$) and dual ($\rho = 20/21$) vortex configuration, where $\delta_{\rm v}$ and $\Delta_{\rm b}$ are defined in Fig.~\ref{IDOSschem}. Low-energy tight-binding models are valid only when $\delta_{\rm v}/\Delta_{\rm b} \ll 1$ as described in the text. We arbitrarily draw the dotted line at $\delta_{\rm v}/\Delta_{\rm b} = 1/2$ to visually indicate the general regions where these tight-binding models are most applicable (i.e., hatched regions indicate where they are not applicable).}
    \label{ratiosfigure}
\end{figure}

For the dual vortex configurations, the low-energy vortex bandwidth $\delta_{\rm v}$ may be estimated as the energy level splitting $2\epsilon$ of the two Majorana modes bound to an isolated white plaquette in a vortex-full sea. Figure 6 of Ref.~\cite{Fuchs20} shows $2\epsilon$ to always be less than the vortex-full background's energy gap $\Delta_{\rm b}$, except near $\kappa = 0$ and $\kappa = 1/2$, where $2\epsilon$ and $\Delta_{\rm b}$ vanish. Thus the $\delta_{\rm v}\ll\Delta_{\rm b}$ condition is essentially equivalent to guaranteeing that $\kappa\neq0$ and $\kappa\neq 1/2$. Dual vortex configurations that are ``dilute'' actually mean that the vortex density $\rho=(n-1)/n \to 1$. 

In Fig.~\ref{ratiosfigure}, we plot the ratio $\delta_{\rm v}/\Delta_{\rm b}$ for the direct $\rho = 1/21$ and dual odd $\rho = 20/21$ vortex configurations as a function of $\kappa$. We offer these two cases as representative examples, as other choices of direct and dual configuration densities exhibit the same qualitative features. As a visual guide, we draw a dotted line at $\delta_{\rm v}/\Delta_{\rm b} = 1/2$ to indicate the regions where $\delta_{\rm v}/\Delta_{\rm b}\ll 1$. For the direct configuration in Fig.~\ref{ratiosfigure}(a), we see that the ratio $\delta_{\rm v}/\Delta_{\rm b}$ is equal to $1/2$ for the two critical values $\kappa_{c_{1}}$ and $\kappa_{c_{2}}$, represented by the red circle and triangle, respectively. In Table~\ref{tablebackgrounds}, we see that the energy gap of the vortex-free ($\rho = 0$) background (roughly equal to $\Delta_{\rm b}$) vanishes as $\kappa = 0$ and is of order $1$ as $\kappa\to\infty$. Thus the low-energy vortex bands are only well-separated from the background as $\kappa_{c_{1}} < \kappa < \kappa_{c_{2}}$. For less dense triangular vortex configurations in the $\rho\to 0$ limit, we find that this region of validity increases, such that $\kappa_{c_{1}}\to 0$ and  $\kappa_{c_{2}}\to \infty$.

For the dual configuration in Fig.~\ref{ratiosfigure}(b), $\delta_{\rm v}/\Delta_{\rm b}$ equals $1/2$ for the three critical values $\kappa_{c_{i}}$ such that $i=1,2,3$. These values are represented by the red circle, triangle, and diamond, respectively. From Sec.~\ref{Sec2} we know that the energy spectrum of the vortex-full background ($\rho = 1$) is gapless as $\kappa=0$ and $\kappa = 1/2$. This establishes two regions of validity for any low-energy effective theory that we wish to build: a small region of low $\kappa$ values such that $\kappa_{c_{1}} < \kappa < \kappa_{c_{2}}$, and a bigger region of large $\kappa$ values such that $\kappa > \kappa_{c_{3}}$. For denser dual configurations in the $\rho\to 1$ limit, we find that these two regions of validity increase, such that $\kappa_{c_{1}}\to 0$, $\kappa_{c_{2}}\to 1/2^{-}$, and $\kappa_{c_{3}}\to 1/2^{+}$.

\section{Dispersion relations of vortex bands}
\label{app:2}
In this Appendix, we analyze three particular cases of quantum phase transitions occurring in the dual odd family of vortex configurations. These transitions are identified by the number of contact points present within the first Brillouin zone, exhibiting one, two, or six such points. The aim is here to show that the effective model captures not only the integrated density of states but really the $\bm{k}$-resolved dispersion relation (at least qualitatively).

\subsection{One contact point}
\begin{figure}
\includegraphics[width=\columnwidth]{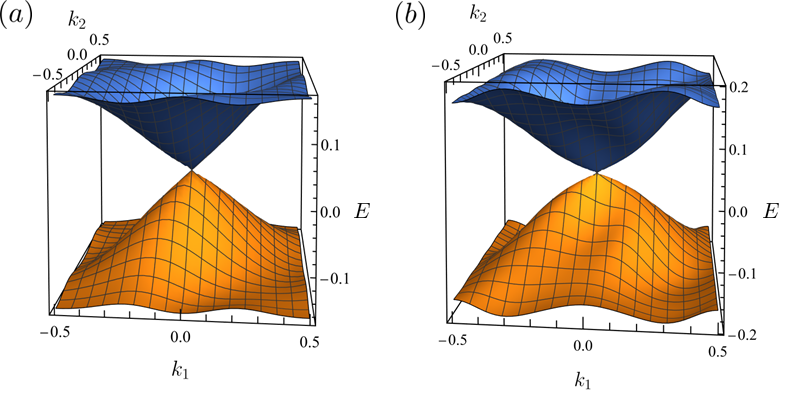}
\caption{(a) Dispersion of the low-lying vortex bands of the Kitaev honeycomb model for the $\rho = 12/13$ dual odd vortex configuration as $\kappa_c = 0.219461$. At this point, the vortex Chern number changes from $\nu_{\rm v} = -1$ to $\nu_{\rm v} = 0$. (b) Dispersion of the effective model of Eq.~\eqref{sablierg}. The parameter values are given by $\epsilon = 0.124$, $t_{1} = 0.0186$, $t_{2} = 0$, $t_{\perp 2} = -0.0124$, and $t_{\perp 1} = \epsilon/6$. The Brillouin zone for both plots is given by $\bm{k} = k_{1}\bm{a}_{1}^{*} + k_{2}\bm{a}_{2}^{*}$, where $(\bm{a}_{1}^{*},\bm{a}_{2}^{*})$ are the reciprocal lattice vectors of $(\bm{a}_{1},\bm{a}_{2})$ respectively, and $|k_{1,2}|\leq 1/2$.}
\label{fig:sablier}
\end{figure}

This type of transition occurs when the vortex Chern number changes from $\nu_{\rm v} = -1$ to $\nu_{\rm v} = 0$ for the dual odd vortex configurations of density $\rho = (n-1)/n$ such that $n$ is not a multiple of three. That is, for densities such as $\rho = 12/13$, $18/19$, $24/25$, and etc. These transitions generally occur when $\kappa_c\sim 0.1-0.2$. Before the transition ($\kappa < \kappa_c$), the system features two gapped Dirac points of opposite chirality and is a Chern insulator with $\nu_{\rm v} = -1$. As $\kappa$ increases, the Dirac points move and at a critical $\kappa_c$ they meet at a time-reversal invariant momentum (TRIM) where they annihilate and the gap closes. Exactly at the transition, there is a single contact point that does not carry a topological charge. After this merging transition, the gap reopens and the system becomes a trivial insulator with $\nu_{\rm v} = 0$. This merging transition is similar to the one occurring in the Kitaev honeycomb model when evolving from the isotropic point $J_{x} = J_{y} = J_{z}$ to a very anisotropic case~\cite{Kitaev06}, or also in uniaxially strained graphene~\cite{Montambaux2009}. 

The Hamiltonian has the same form as that given in Eq.~\eqref{ftandgt}, except that the function $\widetilde{g}(\bm{k})$ is modified. Focusing on the $\rho = 12/13$ vortex density as a example, the ``one contact point" Hamiltonian may be expressed as:
\begin{eqnarray}
\widetilde{g}_{1}(\bm{k}) &=& - i2t_{\perp 1}[\cos(\bm{k}\cdot\bm{a}_{1}) + \cos(\bm{k}\cdot\bm{a}_{2}) + \cos(\bm{k}\cdot\bm{a}_{3})] \nonumber \\ 
&-&2t_{\perp 2}[\sin(\bm{k}\cdot\bm{b}_{1}) + \sin(\bm{k}\cdot\bm{b}_{2}) + \sin(\bm{k}\cdot\bm{b}_{3})] \nonumber\\
&+&i\epsilon.
    \label{sablierg}
\end{eqnarray}
We note that this Hamiltonian results from a different gauge choice for the $s_{jk}^{(\alpha)}$, $s_{jk}^{(\perp)}$, and $\widetilde{s}_{jk}^{(\perp)}$ variables than what is shown in Fig.~\ref{model23}. In Fig.~\ref{fig:sablier}, we plot the dispersion of both this effective model and the low-lying vortex bands of the Kitaev honeycomb model of the $\nu_{\rm v} = -1 \to  0$ transition in the $\rho = 12/13$ dual odd vortex configuration as $\kappa_c = 0.219461$~\cite{Fuchs20}. The proper parameter values for the effective model are given in the caption. The critical point of the effective model appears when $t_{\perp 1} = \epsilon/6$. As $t_{\perp 1} < \epsilon/6$ the Chern number of the effective model is $\nu_{\rm v} = -1$, and when $t_{\perp 1} > \epsilon/6$ the Chern number is $\nu_{\rm v} = 0$. The sign of $t_{\perp 2}$ controls the sign of $\nu_{\rm v}$, and the choice of signs of the cosine terms of $\widetilde{g}_{1}(\bm{k})$ controls the position of the contact point within the Brillouin zone. If we were to instead analyze a different vortex density exhibiting a one contact point transition, one may have to alter the signs of the cosine terms in order to properly match the location of the contact point within the Brillouin zone (this is true for the two and six contact points models as well). Note that because the value of $\delta_{\rm v}/\Delta_{\rm b}\simeq 0.38$ is relatively high (i.e., the vortex bands are not so well separated from the background bands), the bandwidth of the effective model overestimates that of the low-lying vortex bands of the Kitaev honeycomb model.

\subsection{Two contact points}
\begin{figure}
\includegraphics[width=\columnwidth]{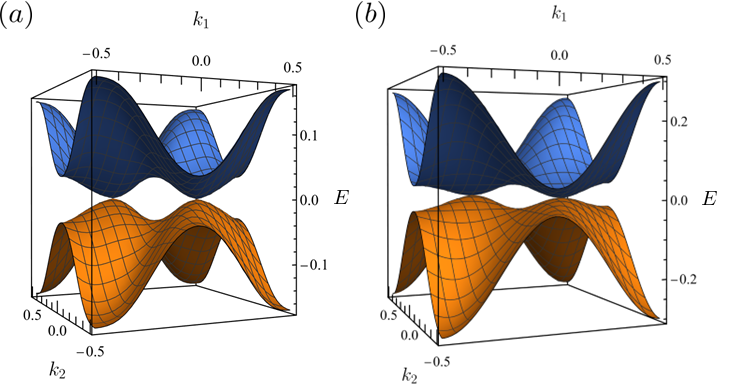}
\caption{(a) Dispersion of the low-lying vortex bands of the Kitaev honeycomb model for the $\rho = 20/21$ dual odd vortex configuration as $\kappa_c = 0.106486$. At this point, the vortex Chern number changes from $\nu_{\rm v} = -2$ to $\nu_{\rm v} = 0$. (b) Dispersion of the effective model of Eq.~\eqref{bicorneg}. The parameter values are given by $\epsilon = 0.1$, $t_{1} = 0$, $t_{2} = t_{\perp 2} = 0.0009$, and $t_{\perp 1} = -\epsilon/3$. The Brillouin zone for both plots is given by $\bm{k} = k_{1}\bm{a}_{1}^{*} + k_{2}\bm{a}_{2}^{*}$ with $|k_{1,2}|\leq 1/2$.}
\label{fig:bicorne}
\end{figure}

This type of transition results in a change of Chern number from $\nu_{\rm v} = -2$ to $\nu_{\rm v} = 0$, and occurs when $\kappa_c\sim 0.1-0.2$ for the dual odd vortex configurations which have a multiple of three in the denominator of their vortex density. That is, for $\rho = 8/9$, $20/21$, $26/27$, and etc. Before the transition, there are four gapped Dirac cones (two of positive and two of negative chirality) and the system is a Chern insulator with 
$\nu_{\rm v} = -2$. At the transition, Dirac points of opposite chirality meet pairwise, merge, and the gap closes. Exactly at the transition, there are two contact points that are not protected by a topological charge and which do not occur at a TRIM. This is possible since this transition may be described as two merging transitions occurring simultaneously. After the transition, a merging gap opens and the system becomes a trivial insulator with $\nu_{\rm v} = 0$.

The Hamiltonian is similar to that of Eq.~\eqref{ftandgt} except for a modified $\widetilde{g}(\bm{k})$ term. Focusing on the $\rho = 20/21$ case as an example, the ``two contact points " Hamiltonian is given by:
\begin{eqnarray}
    \widetilde{g}_{2}(\bm{k}) &=&  i 2t_{\perp 1}[\cos(\bm{k}\cdot\bm{a}_{1}) + \cos(\bm{k}\cdot\bm{a}_{2}) - \cos(\bm{k}\cdot\bm{a}_{3})]\nonumber \\
&-& 2t_{\perp 2}[\sin(\bm{k}\cdot\bm{b}_{1}) - \sin(\bm{k}\cdot\bm{b}_{2}) + \sin(\bm{k}\cdot\bm{b}_{3})]\nonumber \\
&+&i\epsilon.
    \label{bicorneg}
\end{eqnarray}
\begin{figure}
\includegraphics[width=\columnwidth]{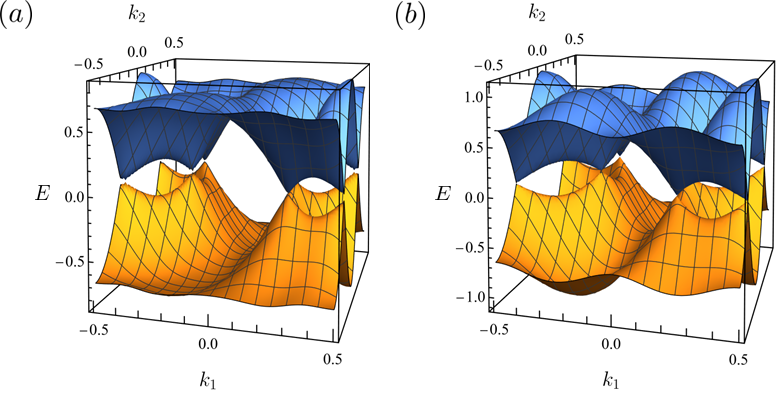}
\caption{(a) Dispersion of the low-lying vortex bands of the Kitaev honeycomb model for the $\rho = 6/7$ dual odd vortex configuration as $\kappa_c = 1.80057$. At this point, the vortex Chern number changes from $\nu_{\rm v} = -3$ to $\nu_{\rm v} = 3$. (b) Dispersion of the effective model of Eq.~\eqref{sixpointg}. The parameter values are given by $\epsilon = t_{\perp 1} = 0.22$, $t_{1} = 0.11$, and $t_{\perp 2} = t_{2} = 0$. The Brillouin zone for both plots is given by $\bm{k} = k_{1}\bm{a}_{1}^{*} + k_{2}\bm{a}_{2}^{*}$ with $|k_{1,2}|\leq 1/2$.}
\label{fig:6diraccones}
\end{figure}
In Fig.~\ref{fig:bicorne}, we plot the dispersion of both this effective model and the low-lying vortex bands of the Kitaev honeycomb model of the $\nu_{\rm v} = -2 \to  0$ transition in the $\rho = 20/21$ dual odd vortex configuration, as $\kappa_c = 0.106486$. The proper parameter values for the effective model are given in the caption. The critical points of the effective model appear when $t_{\perp 1} = -\epsilon/3$. As $t_{\perp 1} > -\epsilon/3$ the Chern number of the effective model is $\nu_{\rm v} = 0$, and when $t_{\perp 1} < -\epsilon/3$ the Chern number is $\nu_{\rm v} = -2$. The sign of $t_{2}$ controls the sign of $\nu_{\rm v}$. Note that $\delta_{\rm v}/\Delta_{\rm b}\simeq 0.62$, and hence the bandwidth of the effective model is slightly larger than that of the low-lying vortex bands of the Kitaev honeycomb model.

We note that in this system, the energy spectrum is gapped and topologically nontrivial only if the sign of one of the sine terms in $\widetilde{g}_{2}(\bm{k})$ is dissimilar to the sign of the other two. If every sine term were positive, for example, then the $2\times 2$ matrix Hamiltonian may be expressed as the linear combination of only two Pauli matrices and would either be gapless before the transition (i.e., a topological semimetal with gapless Dirac points) or gapped after the transition with a trivial Chern number of $\nu_{\rm v} = 0$, due to a merging transition.

\subsection{Six contact points}
This last type of transition results in a change of Chern number from $\nu_{\rm v} = -3$ to $\nu_{\rm v} = 3$, and only occurs for the $\rho = 6/7$ dual odd vortex configuration as $\kappa_c\simeq 1.8$. In contrast to the two previous scenarios, it is not a merging transition of Dirac points, but six Dirac points that each have a simultaneous change in sign of their mass (or gap). The Hamiltonian is similar in form to the previous two except for the $\widetilde{g}(\bm{k})$ term:
\begin{eqnarray}
\widetilde{g}_{6}(\bm{k}) &=& i 2t_{\perp 1}[\cos(\bm{k}\cdot\bm{a}_{1}) - \cos(\bm{k}\cdot\bm{a}_{2}) + \cos(\bm{k}\cdot\bm{a}_{3})]\nonumber \\ 
&-&2t_{\perp 2}[\sin(\bm{k}\cdot\bm{b}_{1}) + \sin(\bm{k}\cdot\bm{b}_{2}) + \sin(\bm{k}\cdot\bm{b}_{3})] \nonumber \\ 
&+& i\epsilon.
\label{sixpointg}
\end{eqnarray}
In Fig.~\ref{fig:6diraccones}, we plot the dispersion of both this effective model and the low-lying vortex bands of the Kitaev honeycomb model of the $\nu_{\rm v} = -3 \to  3$ transition in the $\rho = 6/7$ dual odd vortex configuration, as $\kappa_c = 1.80057$. The proper parameter values for the effective model are given in the caption. The critical points of the effective model appear when $t_{\perp 2} = 0$. As $t_{\perp 2} > 0$ the Chern number of the effective model is $\nu_{\rm v} = -3$, and when $t_{\perp 2} < 0$ the Chern number is $\nu_{\rm v} = 3$. The ratio $\delta_{\rm v}/\Delta_{\rm b}\simeq 0.38$ indicates that the bandwidth of the effective model is slightly larger than that of the low-lying vortex bands of the Kitaev honeycomb model. Note that because $t_{1}\neq t_{2} = 0$, we do not have the same issue of the sine's signs as discussed in the two contact points section, and the $2\times 2$ matrix Hamiltonian is still composed of three different Pauli matrices.

%\bibliography{bibfile}


\end{document}